\documentclass[a4paper,fleqn]{cas-dc}

\usepackage[authoryear,longnamesfirst]{natbib}
\usepackage[nolist]{acronym}
\usepackage{graphicx}
\usepackage{subcaption}
\usepackage{xcolor}
\usepackage{todonotes}
\usepackage{amsthm}

\theoremstyle{remark}
\newtheorem{remark}{Remark}

\usepackage{algorithmic}
\usepackage{algorithm}

\def\tsc#1{\csdef{#1}{\textsc{\lowercase{#1}}\xspace}}
\tsc{WGM}
\tsc{QE}

\usepackage[normalem]{ulem}
\usepackage{soul}
\definecolor{nicegreen}{rgb}{0.0, 0.5, 0.0}

\newcommand{\PS}[1]{\textcolor{black}{#1}}
\newcommand{\PSS}[1]{\textcolor{black}{#1}}
\newcommand{\PSSS}[1]{\textcolor{black}{#1}}

\newcommand{\SAM}[1]{\textcolor{black}{#1}}
\newcommand{\JW}[1]{\textcolor{black}{#1}}

\newcommand{\degC}{\,^{\circ}\mathrm{C}}
\newcommand{\dotdegC}{\cdot^{\circ}\mathrm{C}}
\begin{document}
\let\WriteBookmarks\relax
\def\floatpagepagefraction{1}
\def\textpagefraction{.001}

\shorttitle{}    

\shortauthors{}  

\title [mode = title]{
Peridynamic Thermo-Mechanical Modeling of Cracking in Tungsten under Fusion-Relevant Heat Fluxes}  

\tnotemark[1] 

\tnotetext[1]{
This manuscript has been authored by UT-Battelle, LLC, under contract DE-AC05-00OR22725 with the US Department of Energy (DOE). The US government retains and the publisher, by accepting the article for publication, acknowledges that the US government retains a nonexclusive, paid-up, irrevocable, worldwide license to publish or reproduce the published form of this manuscript, or allow others to do so, for US government purposes. DOE will provide public access to these results of federally sponsored research in accordance with the DOE Public Access Plan (https://www.energy.gov/doe-public-access-plan).
}

\author[1]{Rinkle Juneja}[orcid=0000-0001-6372-1524]
\cormark[1]
\fnmark[1]
\ead{junejar@ornl.gov}
\credit{Conceptualization, Investigation, Writing - Original Draft, \PS{Project administration, Funding acquisition}}

\author[1]{Pablo Seleson}[orcid=0000-0003-3279-4231]
\fnmark[1]
\ead{selesonpd@ornl.gov}
\credit{Conceptualization, Methodology, \PS{Software},  Validation, \PS{Formal analysis}, Investigation,  Writing - Original Draft\PS{, Visualization}}

\author[1]{Samuel Temple Reeve}[orcid=0000-0002-4250-9476]
\fnmark[1]
\ead{reevest@ornl.gov}
\credit{Conceptualization, \PS{Methodology,} Software,  Investigation, \PS{Data Curation}, Writing - Original Draft, \PS{Visualization}}
\author[1]{Justin Weinmeister}[orcid=0000-0001-6009-0237]
\ead{weinmeistejr@ornl.gov}
\credit{\PS{Validation,} Investigation, Writing - Review \& Editing}
\author[1]{Gary Staebler}[orcid=0000-0002-1944-1733]
\ead{staeblergm@ornl.gov}
\credit{\PS{Writing - Review \& Editing, Project administration,} Funding acquisition}

\affiliation[1]{organization={Oak Ridge National Laboratory},
            city={Oak Ridge},
            postcode={37830}, 
            state={TN},
            country={USA}}

\cortext[1]{Corresponding author}

\fntext[1]{Contributed equally.}

\begin{abstract} \glsresetall
\Acp{PFM} are the primary armor materials in  fusion reactors, protecting reactor wall components from extreme heat fluxes, energetic particles, and intense neutron irradiation. These harsh conditions induce thermo-mechanical stresses, irradiation damage, and overall degradation in \acp{PFM}. Among these, thermo-mechanical damage induced by extreme heat flux is one of the key factors limiting the lifetime of \acp{PFM}. 
We present a modeling and simulation approach for predicting thermo-mechanical cracking \PSSS{and damage} in tungsten, a leading \ac{PFM} candidate, under fusion-reactor-relevant heat fluxes. 
Specifically, we study damage induced by cyclic electron beam (e-beam) exposure under different combinations of base temperature and heat flux. 
We carry out component-scale thermo-mechanical  simulations using CabanaPD, a peridynamics fracture mechanics code, informed by experimental data for recrystallized pure tungsten from the JUDITH e-beam facility. The simulations compare well with the experimental classification and 
enable identification of 
\PSSS{cracking} 
thresholds across a range of heat fluxes and base temperatures. These predictions can provide detailed insights into the spatial distribution and temporal accumulation of cracks and damage, 
while offering a pathway for screening candidate \acp{PFM} and quantifying their lifespan under fusion-relevant conditions. 
\end{abstract}

\begin{acronym}
  \acro{PFM}{plasma-facing material}
  \acro{PD}{peridynamics}
  \acro{DBTT}{ductile-to-brittle transition temperature}
  \acro{IGP}{ITER Grade Plansee}
  \acro{ADR}{adaptive dynamic relaxation}
  \acro{EPP}{elastic-perfectly plastic}
  \acro{PMB}{prototype microelastic brittle}
  \acro{CTE}{coefficient of linear thermal expansion}
\end{acronym}


\maketitle

\section{Introduction}

The realization of fusion energy 
critically depends on the reliability of materials under the extreme conditions of a fusion reactor. 
\Acfp{PFM} are exposed to steady-state heat fluxes of the order of~10-20 MW/m$^2$ while repetitive transient heat fluxes can reach up to $\sim$1 GW/m$^2$~\citep{linke2019, Linsmeier_2017}. 
Tungsten remains one of the most promising metallic \acp{PFM} due to its high melting point and high thermal conductivity, among other \PS{properties required for} long-term plasma exposure.
However, a chief limitation of tungsten is its susceptibility to cracking and failure under repeated thermal shocks~\citep{Loewenhoff_2011}. Developing \PS{modeling and simulation capabilities} that accurately predict material damage and fracture is therefore critical for optimizing the performance and extending the lifetime of \acp{PFM}.

Multiple research efforts have focused on predicting thermally-driven crack propagation in \acp{PFM} at the continuum scale 
\citep{li2015, chen2025, Changjun2017, Durif2023}.
However, many issues \PS{have limited} the impact of these simulations to date. The short \PS{time scales} of individual thermal pulses\PS{, together with the large number of} pulses necessary to cause significant damage\PS{,} make simulations computationally expensive, while the state\PS{-} and cycle-dependent material properties require detailed model\PS{ing}. While \PS{beyond the scope} of this work, \PS{the} complexity increases substantially when modeling material degradation and failure \PS{due to both thermal cycling and plasma irradiation}~\citep{Durif2022}. %
In addition, existing methods \PS{are largely based on classical continuum mechanics---typically implemented using the finite element method (FEM)---whose governing partial differential equations require smooth fields. Consequently, these methods} are not naturally suited to \PS{simulating cracking and other discontinuities and often require special treatments, such as remeshing~\citep{SwensonIngraffea1988}, interelement separation governed by cohesive laws~\citep{XuNeedleman1994}, or enrichment of the
FEM polynomial approximation space to capture field 
discontinuities within elements~\citep{BelytschkoBlack1999, MoësDolbowBelytschko-1999}.} Finally, \PS{the} limited availability of systematic data under reactor-relevant loading conditions continues to make model development and validation difficult. 

\PS{In this work, we develop a \ac{PFM}-specific continuum-scale model based on peridynamics (PD) and apply it to predict cyclic cracking and damage in recrystallized pure tungsten under fusion-relevant heat fluxes. 
\PSSS{PD} 
employs an integro-differential formulation that, unlike classical continuum mechanics, \PSSS{does not require spatial differentiability of the field variables and is therefore particularly}
well suited to modeling cracking and damage.}   
\PS{The model employs a fully coupled thermo-mechanical and heat-transfer formulation, with parameters derived from the material properties reported in~\citep{wirtz2013} and~\citep{wirtz2016}, which together constitute a consistent and nearly complete experimental dataset. Using the thermal process conditions reported in the aforementioned works, %
we simulate damage induced by cyclic electron beam (e-beam) exposure over a range of base temperatures and heat fluxes. Despite the continuum-level assumptions, the simulations show good agreement with experiments and represent an important step toward quantitative \ac{PFM} lifetime prediction.}

The rest of this paper is organized as follows. Section~\ref{sec::pd} describes the methodological approach and the \ac{PD} modeling framework used to predict thermally induced cracking and damage. Section~\ref{sec::param} presents the experimentally measured material properties necessary for the simulations. Section~\ref{sec:sim} provides the e-beam conditions and simulation details. Section~\ref{sec::results} reports the simulation results and compares them with experiments. Finally, Section~\ref{sec::conclusions} provides a summary of the study and outlines future extensions.

\section{Methodology}
\label{sec::pd}

Thermo-mechanical modeling was carried out using \ac{PD},  a nonlocal reformulation of classical continuum mechanics tailored to the simulation of material failure and damage~\citep{Silling2000, Silling2007}. 
This approach is particularly powerful for capturing complex \PS{fracture} phenomena, including the nucleation, propagation, and branching of cracks during dynamic loading~\citep{Ha2010,Silling2010}. Most notably,  \ac{PD} has been shown to successfully capture the formation and propagation of complex cracks induced by thermal shocks \PS{in} comparison to experiment~\citep{GIANNAKEAS20183037, DANTUONO201774}.

For the \ac{PD} simulations presented in this work, the CabanaPD code was employed~\citep{cabanapd}\footnote{https://github.com/ORNL/CabanaPD} with fully-coupled thermal and mechanical models (see Section~\ref{sec: Peridynamic thermo-mechanical modeling}). CabanaPD is a performance-portable \ac{PD} code designed to run on leadership-class supercomputers as well as local workstations. CabanaPD uses the Kokkos library for performance\PS{-}portable threaded parallelism across hardware~\citep{kokkos} and the Cabana library for portable and distributed particle simulation~\citep{cabana}, \PS{providing} a meshfree implementation of \ac{PD}~\citep{SILLING2005}. Together, Kokkos and Cabana enable CabanaPD to scale effectively across many high\PS{-}performance computing systems~\citep{CoPA-paper}. 
While CabanaPD supports a range of \ac{PD} models with varying levels of complexity, we focus \PS{here} on bond-based \ac{PD} models to balance accuracy and computational efficiency. 

\subsection{Peridynamic thermo-mechanical modeling}
\label{sec: Peridynamic thermo-mechanical modeling}

In PD, the fundamental quantity is the bond, 
defined as the vector connecting two material points in the reference configuration. Each material point is assumed to interact with neighboring material points through bonds, forming a nonlocal modeling framework. In bond-based \ac{PD} models, the deformation of each bond is independent of other bonds, thereby reducing computational complexity. In this study, fully-coupled thermal and mechanical models are employed. 
The mechanics model incorporates bond failure, plastic deformation, and thermal expansion. The multiphysics  coupling is bidirectional: the heat transfer model informs the mechanics model about temperature changes \PS{that induce} thermal expansion, while the mechanics model informs the heat transfer model about fracture events that generate discontinuities in the thermal diffusion pathways. 

\subsubsection{Mechanics}
\label{sec: mechanics}

The mechanical response is governed by a bond-based \ac{PD} formulation~\citep{Silling2000}. Given a bounded body $\mathcal{B}\subset \mathbb{R}^3$, the equation of motion for a material point ${\bf x}\in \mathcal{B}$ at time $t\geqslant 0$, in the absence of external body forces, is 
\begin{equation}
\label{eq: PD mechanics}
\rho(T_0) 
\frac{\partial^2 {\bf u}}{\partial t^2}({\bf x},t) 
= \int_{\mathcal{H}_{\bf x}} {\bf f}({\bf x}^\prime, {\bf x},t) dV_{\bf x^\prime},
\end{equation}
where $\mathbf{x}$ denotes the position in the reference configuration, $\rho$ is the mass density (assumed in this work to be temperature-dependent and given by \eqref{eq: density}), $T_0$ is the reference temperature, 
${\bf u}$ is the displacement field, ${\bf f}$ is a pairwise force function, 
\PS{$dV_{{\bf x}^\prime}$ is the differential volume element associated with the neighboring material point ${\bf x}^\prime$},
and $\mathcal{H}_{\bf x}$ is the nonlocal neighborhood of ${\bf x}$ defined by
\begin{equation*}
\mathcal{H}_{\bf x}:=\{ {\bf x}^\prime \in \mathcal{B}: \|{\bf x}^\prime -{\bf x}\|\leqslant \delta\},
\end{equation*}
where $\|\cdot\|$ denotes the Euclidean norm \PS{and} $\delta>0$ is the maximum distance over which nonlocal interactions occur (called the horizon)\PS{.} %

The pairwise force function describes the mechanical response of each bond. 
In this work, we employ a thermo-elastoplastic formulation with perfect plasticity and temperature-dependent mechanical properties (see Section~\ref{sec::param}). The constitutive relation between mechanical forces and deformation is given by the following pairwise force function:
\begin{equation}
\label{eq: pairwise force function}
{\bf f} = \mu \frac{c + c^\prime}{2} \left(
s - s_{\rm th} - s_{\rm p}\right) \frac{{\bf y}^\prime- {\bf y}}{\|{\bf y}^\prime- {\bf y}\|},
\end{equation}
where $\mu$ is a bond-breaking function; $c$ and $c^\prime$ are the elastic \PS{micromodulus} constants \PS{associated with} ${\bf x}$ and ${\bf x}^\prime$, respectively\PS{, where the elastic micromodulus is a \ac{PD} constitutive parameter governing pairwise mechanical interactions;} 
$s$, $s_{\rm th}$, and $s_{\rm p}$ are the total, thermal, and plastic stretches of a bond, respectively;  
and ${\bf y}^\prime$ and ${\bf y}$ denote the positions of  ${\bf x}^\prime$ and ${\bf x}$, respectively, in the current configuration. Note that the total stretch is composed of  elastic, thermal, and plastic contributions, and the elastic  stretch can be expressed as 
\begin{equation}
\label{eq: elastic stretch}
    s_{\rm e} = s - s_{\rm th} - s_{\rm p}.
\end{equation}
In fact, the pairwise force function in \eqref{eq: pairwise force function}  depends on the deformation through the elastic  stretch.  
For brevity, the arguments are omitted in~\eqref{eq: pairwise force function} and~\eqref{eq: elastic stretch}; the variable dependencies are specified below. For convenience, we use  the notation $T$ and $T^\prime$ below to denote the temperatures at ${\bf x}$ and ${\bf x}^\prime$ at any given time. 

\begin{remark}
Although this work concerns homogeneous materials, the temperature dependence of the material parameters requires that the material constants at both material points connected by a bond 
be taken into account, as is done for the elastic \PS{micromodulus} constants in~\eqref{eq: pairwise force function}. In this work, 
the arithmetic average of the corresponding material parameters is adopted.
\end{remark}

The elastic \PS{micromodulus} constant 
is taken from the \ac{PMB} material model~\citep{SILLING2005}, with temperature dependence incorporated, and is given by
\begin{equation}
\label{eq: micromodulus constant}
c(T) = \frac{6E(T)}{\pi \delta^4 (1-2\nu)},
\end{equation}
where $E$ is the temperature-dependent Young's modulus (given in this work by~\eqref{eq: Young's modulus}) and $\nu$ is the Poisson's ratio\PS{. In bond-based PD, $\nu$ is not an independent material parameter because the pairwise central-force assumption admits only a single constitutive parameter for isotropic linear elasticity~\citep{Silling2000, Trageser2020}; therefore, in three dimensions, 
$\nu = 0.25$.} %
Note that in~\eqref{eq: pairwise force function}, $c = c(T)$ and $c^\prime = c(T^\prime)$. 

The total stretch %
represents the engineering strain in a bond and is given by ~\citep{SILLING2005}
\begin{equation*}
s({\bf x}^\prime, {\bf x},t) 
 = \frac{\|{\bf y}({\bf x}^\prime,t)- {\bf y}({\bf x},t) \|- \|{\bf x}^\prime- {\bf x} \|}{\|{\bf x}^\prime- {\bf x} \|}.
\end{equation*}
Note that in~\eqref{eq: pairwise force function}, ${\bf y}={\bf y}({\bf x},t) = {\bf x} + {\bf u}({\bf x},t)$ and ${\bf y}^\prime ={\bf y}({\bf x}^\prime,t) = {\bf x}^\prime + {\bf u}({\bf x}^\prime,t)$. 
The thermal stretch 
is given by
\begin{equation}
\label{eq: thermal stretch}
s_{\rm th}(T^\prime,T) = 
\frac{\alpha(T)+\alpha(T^\prime)}{2}
\left(
\frac{T+T^\prime}{2}-T_0
\right),
\end{equation}
where $\alpha$ is 
the temperature-dependent \ac{CTE} (given in this work by~\eqref{eq: CTE}).  
In contrast to the total and thermal stretches, the plastic stretch, $s_{\rm p}$, is treated as a history-dependent internal variable associated with each bond.
In this work, perfect plasticity is assumed, and $s_{\rm p}$ is evolved by adapting the 
microplastic model of~\cite{MACEK2007} to include thermal deformation and a temperature-dependent yield stretch, \PS{$s_{y}$}. Initially, $s_{\rm p} =0$ at $t=0$. Under the elastic--perfectly plastic assumption, $s_{\rm p}$ evolves so as to enforce the yield constraint $|s_{\rm e}|\leqslant \PS{s_{y}}$. 
When $|s_{\rm e}|< \PS{s_{y}}$, the response is elastic and $s_{\rm p}$ remains unchanged. Once the yield condition is reached, any further deformation  
is accommodated through  the evolution of $s_{\rm p}$ via a return-mapping procedure, such that $s_{\rm e}=\PS{s_{y}}$ in tension and $s_{\rm e}=\PS{-s_{y}}$ in compression. To this end, a trial elastic stretch, $s_{\rm e}^{\rm trial}$, is defined and compared with \PS{$s_{y}$}, as shown in \PS{Algorithm}~\ref{alg: Update s_p}. 
\PS{In the present model, $s_{y}$ is defined directly from the yield strain as}
\begin{equation}
\label{eq: yield stretch}
\PS{s_{y}}(T^\prime,T) = \frac{\PS{\varepsilon_{y}} (T) + \PS{\varepsilon_{y}} (T^\prime)}{2},
\end{equation}
where \PS{$\varepsilon_{y}$} 
is the temperature-dependent yield strain
(given in this work by~\eqref{eq: yield strain})\PS{; this definition is consistent with the approach adopted below for the critical stretch.}
\begin{algorithm}
\caption{: Update of plastic stretch $s_{\rm p}$ by return mapping}
\label{alg: Update s_p}
\begin{algorithmic}[1]
  \footnotesize
\STATE \COMMENT{Given \PS{$s_{y}$}, $s$, $s_{\rm th}$, and $s_{\rm p}$; $s_{\rm p} = 0$ at $t=0$}
\STATE $s_{\rm e}^{\rm trial}  =  s - s_{\rm th} - s_{\rm p}$
\IF{$s_{\rm e}^{\rm trial}  \geqslant \PS{s_{y}}$}
	\STATE $s_{\rm p} = s - s_{\rm th} \PS{- s_{y}}$ 
    \COMMENT{yielding in tension}
\ELSIF{$s_{\rm e}^{\rm trial} \leqslant \PS{- s_{y}}$}
        \STATE $s_{\rm p} = s - s_{\rm th}  + \PS{s_{y}}$
        \COMMENT{yielding in compression}
\ELSE
	\STATE $s_{\rm p}$ remains unchanged 
    \COMMENT{elastic response}
\ENDIF
\STATE Return $s_{\rm p}$
\end{algorithmic}
\end{algorithm}

The bond-breaking function, $\mu$, is a Boolean-valued, history-dependent function that governs bond failure and thereby enables the description of material damage. Similar to the plastic stretch, $\mu$ is an internal variable associated with each bond. 
Here, we employ a bond-failure criterion by adapting the critical-stretch criterion of~\cite{SILLING2005} to the thermo-elastoplastic formulation. The bond-breaking function 
$\mu$ is initialized to \PS{$1$} and is irreversibly set to \PS{$0$} when %
$s_{\rm e} + s_{\rm p} \geqslant s_0$, 
where $s_0$ is the critical stretch\PS{. In the present model, $s_0$ is defined directly from the failure strain as} %
\begin{equation}
\label{eq: critical stretch}
s_0(T^\prime,T) = \frac{\varepsilon_{\rm fail}(T) + \varepsilon_{\rm fail}(T^\prime)}{2},
\end{equation}
where $\varepsilon_{\rm fail}$ is 
the temperature-dependent failure strain
(determined in this work as described in Section~\ref{sec: Yield strain and failure strain})\PS{; a similar definition was used by~\cite{ZU20242349}, although with a constant critical stretch.} 
The damage at a material point is defined as
\begin{equation}
\label{eq: damage}
\varphi({\bf x},t) := 1 - \frac{\int_{\mathcal{H}_{\bf x}} \mu({\bf x}^\prime, {\bf x},t) dV_{\bf x^\prime}}{\int_{\mathcal{H}_{\bf x}} dV_{\bf x^\prime}}.
\end{equation}
This variable satisfies $0\leqslant \varphi \leqslant 1$ as it is \PS{defined} relative to the \PS{initially} intact bonds. 
A material point with all of its bonds intact has a damage  value of $\varphi=0$, whereas a material point with all of its bonds broken has a damage value of $\varphi=1$.

\subsubsection{Heat transfer}
\label{sec: heat transfer}

Nonlocal heat transfer uses ``thermal bonds,'' analogous to the mechanical concept of \ac{PD} bonds~\citep{BOBARU2010}. 
The governing equation for a material point ${\bf x}\in \mathcal{B}$ at time $t\geqslant 0$, in the absence of volumetric heat sources, is 
\begin{equation}
\label{eq: PD heat transfer}
\rho(T_0) c_p(T({\bf x},t)) 
\frac{\partial T}{\partial t}({\bf x},t)
= \int_{\mathcal{H}_{\bf x}} 
\Theta({\bf x}^\prime, {\bf x},t)
dV_{\bf x^\prime},
\end{equation}
where  $c_p$ is the temperature-dependent specific heat capacity (given in this work by~\eqref{eq: Specific heat capacity}) and $\Theta$ is a pairwise heat transfer function  given by
\begin{equation}
\label{eq: pairwise heat transfer function}
\Theta
= \mu \frac{k + k^\prime}{2} \frac{T^\prime - T }{\|{\bf x}^\prime - {\bf x} \|^2},
\end{equation}
where $k$ and $k^\prime$ are the \PS{thermal microconductivities associated with} ${\bf x}$ and ${\bf x}^\prime$, respectively\PS{, where the thermal microconductivity is a \ac{PD} constitutive parameter governing pairwise heat transfer}. %
As in~\eqref{eq: pairwise force function}, the arguments are omitted in~\eqref{eq: pairwise heat transfer function} for brevity. 

The \PS{thermal microconductivity} is taken as~\citep{BOBARU20122764}
\begin{equation}
\label{eq: micro-conductivity}
k(T) = \frac{9\kappa(T)}{2\pi\delta^3},
\end{equation}
where $\kappa$ is the temperature-dependent thermal conductivity (given in this work by~\eqref{eq: Thermal conductivity}). Note that in~\eqref{eq: pairwise heat transfer function}, $k = k(T)$ and $k^\prime = k(T^\prime)$.

\subsection{Discretization and time stepping}

\subsubsection{Meshfree approach}
\label{sec: Meshfree approach}

In the meshfree approach of~\cite{SILLING2005}, a body $\mathcal{B}$ is discretized into a set of $N$ computational particles with reference positions $\{ {\bf x}_i\}_{i=1}^N$, and each particle~$i$ is assigned a material volume $V_i$. For each particle $i$, its family is defined as
\begin{equation*}
\mathcal{F}_i:=\{j\neq i:\|{\bf x}_j- {\bf x}_i\|\leqslant \delta\},
\end{equation*}
which is the set of indices of all particles within the neighborhood of particle~$i$. 
Using this approach, the discrete mechanics equation for particle $i$ at time $t^n$ \PS{is}  
({\em  cf.}~\eqref{eq: PD mechanics} and~\eqref{eq: pairwise force function})
\begin{equation*}
\rho_0
\ddot{\bf u}_i^n 
= \sum_{j\in \mathcal{F}_i} 
 \mu_{ji}^n \frac{c_i^n + c_j^n}{2} \left(
s_{ji}^n - s^{n}_{{\rm th},ji} - s^{n}_{{\rm p},ji} \right) \frac{{\bf y}_j^n- {\bf y}_i^n}{\|{\bf y}_j^n- {\bf y}_i^n\|}V_j
\end{equation*}
and the discrete heat transfer equation for particle $i$ at time $t^n$ \PS{is}  
({\em cf.}~\eqref{eq: PD heat transfer} and~\eqref{eq: pairwise heat transfer function})
\begin{equation*}
\rho_0
c_{p,i}^n \dot{T}_i^n
 = \sum_{j\in \mathcal{F}_i}
\mu_{ji}^n \frac{k_i^n + k_j^n}{2} 
\frac{T_j^n - T_i^n }{\|{\bf x}_j - {\bf x}_i \|^2}
V_j.
\end{equation*}
Here, subscripts $i$ and $j$ denote evaluation at ${\bf x}_i$ and ${\bf x}_j$, respectively, and the superscript $n$ denotes evaluation at time $t^n$. 
\PS{For brevity, we use the notation\\}
\[
\begin{alignedat}{3}
\rho_0 & := \rho(T_0), \quad & 
\ddot{\bf u}_i^n & : = \frac{\partial^2 {\bf u}}{\partial t^2}({\bf x}_i,t^n), \quad &
\mu_{ji}^n & := \mu({\bf x}_j, {\bf x}_i,t^n),\\
c_i^n &:=c(T_i^n),&
{\bf y}_i^n &:  = {\bf y}({\bf x}_i,t^n),& 
s_{ji}^n & :=s({\bf x}_j, {\bf x}_i,t^n),\\
c_{p,i}^n & :=c_p(T_i^n),&
k_i^n & :=k(T_i^n),&
s^{n}_{{\rm th},ji} & :=s_{\rm th}(T_j^n,T_i^n),\\
T_i^n & :=T({\bf x}_i,t^n), \quad &
\dot{T}_i^n & :=\frac{\partial T}{\partial t}({\bf x}_i,t^n),&
s^{n}_{{\rm p},ji} & :=s_{{\rm p}}({\bf x}_j,{\bf x}_i,t^n).
\end{alignedat}
\]
Similarly, the damage at particle $i$ at time $t^n$ is computed as ({\em cf.}~\eqref{eq: damage})
\begin{equation}
\label{eq: damage discrete}
\varphi_i^n := 1 - \frac{\sum_{j\in \mathcal{F}_i} \mu_{ji}^n V_j}{\sum_{j\in \mathcal{F}_i} V_j},
\end{equation}
where $\varphi_i^n := \varphi({\bf x}_i,t^n)$. Using the particle damage field, we define the total damage at time $t^n$ as
\begin{equation}
\label{eq: total damage discrete}
D^n := \frac{\sum_{i=1}^N \varphi_i^n V_i}{\sum_{i=1}^N V_i}.
\end{equation}

\subsubsection{Time stepping}
\label{sec:Time stepping}
Both explicit and quasi-static time-stepping schemes are available in CabanaPD. 

For mechanics, the default method is the explicit velocity-Verlet scheme with a fixed time step. Following~\cite{SILLING2005}, \PS{and using the expression for the elastic micromodulus constant from~\eqref{eq: micromodulus constant}}, a stable \PS{mechanical} time step, $\Delta t_{\rm m}$, for a given safety factor, $\eta_{\rm m}$, can be written as
\begin{equation}
\label{eq: time step mechanics}
\Delta t_{\rm m} = \eta_{\rm m} \sqrt{\frac{\rho_{0}}{\sum_{j\in \mathcal{F}_i}
\frac{6E_0}{\pi\delta^4\|{\bf x}_j - {\bf x}_i\|}V_j}},
\end{equation}
where $E_0:=E(T_0)$.
Since $E(T)$ decreases monotonically with $T$ (see Figure~\ref{fig: Elastic modulus}), and the temperature does not fall below $T_0$  during the simulation, $E_0$ is the maximum value of the Young's modulus attained in the simulation. 

For heat transfer, the default method is the explicit forward Euler scheme with a fixed time step. As in mechanics, a stable \PS{thermal} time step, $\Delta t_{\rm th}$, for a given safety factor, $\eta_{\rm th}$, can be derived\PS{. Using the expressions for the elastic micromodulus constant and the thermal microconductivity given in~\eqref{eq: micromodulus constant} and~\eqref{eq: micro-conductivity}, respectively, it can be written} as 
\begin{equation}
\label{eq: time step heat trasnfer}
\Delta t_{\rm th} = 
\eta_{\rm th} 
\frac{8\rho_{0}E(T_{\rm max})}{\sum_{j\in \mathcal{F}_i}
\frac{3\kappa_0\delta
}{\|{\bf x}_j - {\bf x}_i\|^2}V_j},
\end{equation}
where $\kappa_0:=\kappa(T_0)$ and $T_{\rm max}$ denotes the maximum temperature attained in the simulation\PS{,} which can be determined from Figure~\ref{fig: T interpolation}.
Similar to $E(T)$, $\kappa(T)$ decreases monotonically with $T$ (see Figure~\ref{fig: kappa}); since the temperature does not fall below $T_0$  during the simulation, $\kappa_0$ is the maximum value of the thermal conductivity attained in the simulation. 

Quasi-static time evolution for mechanics in CabanaPD uses \ac{ADR}, following~\citet{wang2018}. In this work, the \PS{explicit and quasi-static time-stepping} approaches are applied iteratively\PS{.} 
Damage is not allowed to evolve during the quasi-static portion of the simulation.

\section{Experimental material properties}
\label{sec::param}

The parameters of the \ac{PD} thermo-mechanical model are determined from experimentally measured material properties of tungsten reported by~\cite{wirtz2013} and~\cite{wirtz2016}, as described below.  
Here, we focus on recrystallized \ac{IGP} tungsten\PS{, manufactured by Plansee SE, Austria,}
with a purity of 99.97 wt$\%$\PS{, referred to in those works and throughout this paper as ``pure tungsten.'' The} forged specimens were recrystallized at $1600\degC$ for one hour. 
\PS{Table~\ref{tab: material_properties_pd} summarizes the experimentally measured material properties used in the \ac{PD} thermo-mechanical model, together with the 
\PSSS{equations and/or figures in the sections below where they are defined} 
and the \ac{PD} model equations in which they appear. These properties \SAM{include} mechanical properties (Section~\ref{sec: Mechanical properties}), thermal properties (Section~\ref{sec: Thermal properties}), and physical properties (Section~\ref{sec: Physical properties}).
}

\PS{
\begin{table}[h!]
\centering
\begin{tabular}{|p{3.5cm}p{2.2cm}p{1.6cm}|}
 \hline
  {\bf Property} & {\bf Reference(s)} & {\bf PD model}  \\
 \hline 
Young's modulus ($E$) & Eq.~\eqref{eq: Young's modulus}, Fig.~\ref{fig: Elastic modulus} & Eq.~\eqref{eq: micromodulus constant}  \\
Yield strain ($\varepsilon_{y}$) & Eq.~\eqref{eq: yield strain}, Fig.~\ref{fig: Yield strain and failure strain} &  Eq.~\eqref{eq: yield stretch}\\ 
Failure strain ($\varepsilon_{\rm fail}$) & Fig.~\ref{fig: Yield strain and failure strain} & Eq.~\eqref{eq: critical stretch} \\ 
CTE ($\alpha$) & Eq.~\eqref{eq: CTE}, Fig.~\ref{fig: CTE} & Eq.~\eqref{eq: thermal stretch} \\
Specific heat capacity  ($c_p$) & Eq.~\eqref{eq: Specific heat capacity}, Fig.~\ref{fig: c_p} & Eq.~\eqref{eq: PD heat transfer} \\
Thermal conductivity ($\kappa$) & Eq.~\eqref{eq: Thermal conductivity}, Fig.~\ref{fig: kappa} & Eq.~\eqref{eq: micro-conductivity} \\
Mass density ($\rho$) & Eq.~\eqref{eq: density}, Fig.~\ref{fig: Density} & Eqs.~\eqref{eq: PD mechanics}, \eqref{eq: PD heat transfer} \\
 \hline
\end{tabular}
\caption{\PS{Material properties and their connections to the \ac{PD} model. The listed properties correspond to the mechanical, thermal, and physical properties introduced in Sections~\ref{sec: Mechanical properties}--\ref{sec: Physical properties}. Each row presents a material property, its reference(s) from  equations and/or figures, and the corresponding \PSSS{PD} equation(s) in Section~\ref{sec::pd} in which it appears. Some properties also appear in the stable mechanical or thermal time-step equations in Section~\ref{sec:Time stepping}; see \eqref{eq: time step mechanics} and \eqref{eq: time step heat trasnfer}, respectively.}}
\label{tab: material_properties_pd}
\end{table}
}

\subsection{Mechanical properties}
\label{sec: Mechanical properties}

\subsubsection{Young's modulus}
\label{sec: Young's modulus}

\PS{The Young's modulus, $E$, is used in the \ac{PD} model in the equation for the elastic micromodulus constant (see~\eqref{eq: micromodulus constant}). Its temperature dependence was}
reported in~\citet[Figure~3.13]{wirtz2013}, \PS{based on experimental measurements} using the impulse excitation method. The reported \PS{values of the} Young's modulus \PS{were obtained using} a formula that depends on \PS{the} Poisson's ratio, which was taken as $\nu = 0.3$. 
\PS{As discussed earlier, however, we note the bond-based \ac{PD} model used in this work is restricted to $\nu = 0.25$.} %
\PS{The} measurements were carried out in ambient air; therefore, the temperature range was limited to between room temperature and $400\degC$, to avoid issues with oxidation at higher temperatures. 

\begin{figure}[h!]
    \centering
    \includegraphics[width=0.5\textwidth, trim=0cm 0cm 0cm 0cm, clip]{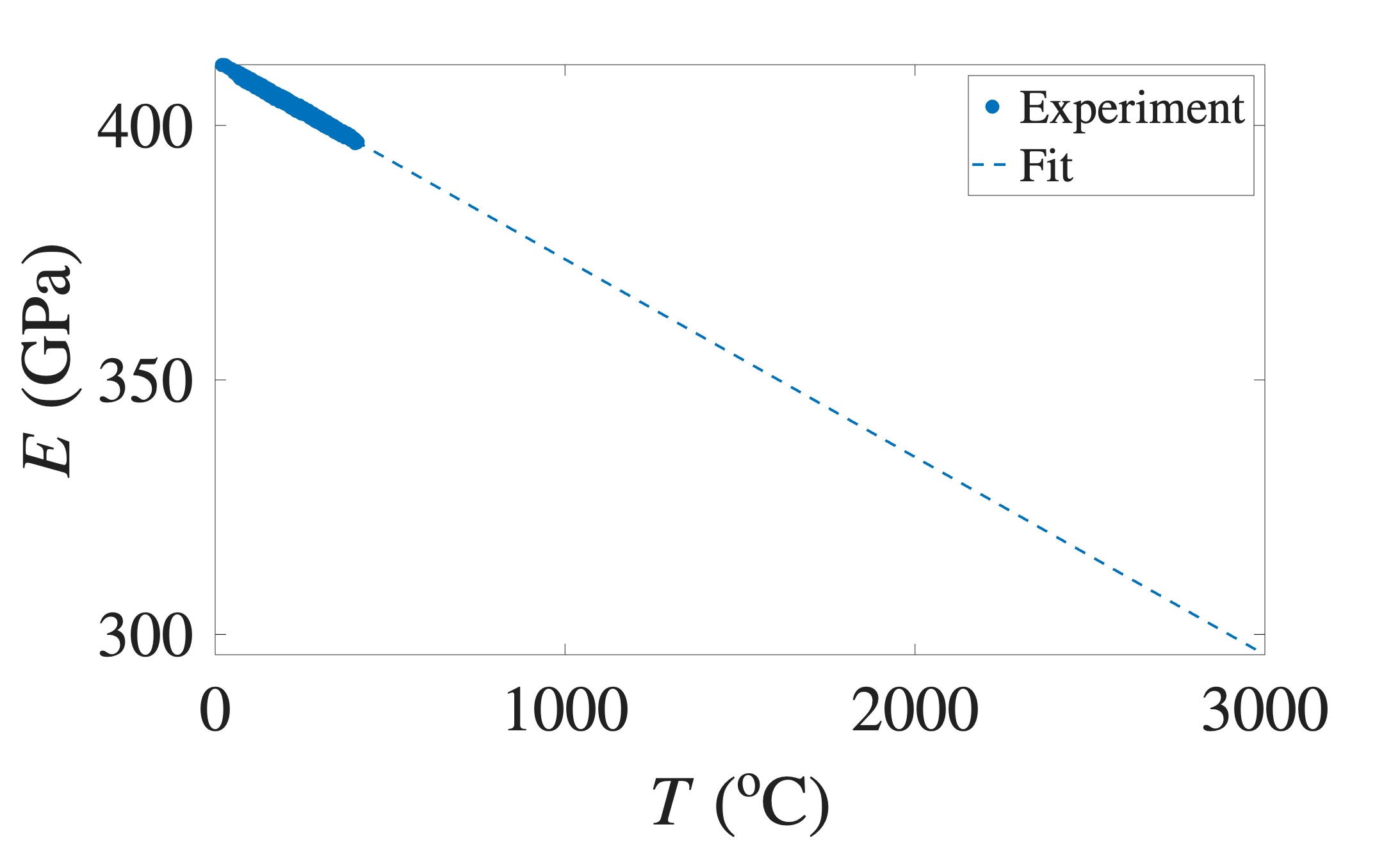}
    \caption{Temperature dependence of the \PS{Young's modulus}. Markers indicate experimental data from~\cite{wirtz2013}; the dashed line indicates the linear fit. %
    }
    \label{fig: Elastic modulus}
\end{figure}

The digitized experimental data and a linear fit are presented in Figure~\ref{fig: Elastic modulus}. 
We note that the base temperatures of interest in this study include values above $400\degC$, and even when base temperatures are lower, temperatures can significantly exceed this threshold during a thermal shock; therefore, we extrapolate the linear fit. 
Figure~\ref{fig: Elastic modulus} shows the  extrapolation up to $3000\degC$, which is approximately the maximum temperature reached for a heat flux of $1.8$ GW/m$^2$ from a base temperature of $23\degC$ (see Figure~\ref{fig: T interpolation}). Starting from a higher base temperature results in even higher peak temperatures. Note that the \PSSS{melting} temperature of tungsten is $3422\degC$~\PS{\citep{wirtz2013}}; however, melting is not considered in the present modeling framework. 
The resulting fitted function, with temperature, $T$, in\PS{$\degC$} and Young's modulus, $E$, in GPa is:
\begin{equation}
\label{eq: Young's modulus}
E(T) = -3.88\times 10^{-2}\, T + 4.13\times 10^{2}.
\end{equation}

\subsubsection{Yield strain and failure strain}
\label{sec: Yield strain and failure strain}

\begin{figure}[h!]
    \centering
    \begin{subfigure}[b]{0.5\textwidth}
        \centering
        \includegraphics[width=\textwidth, trim=0cm 0cm 0cm 0cm, clip]{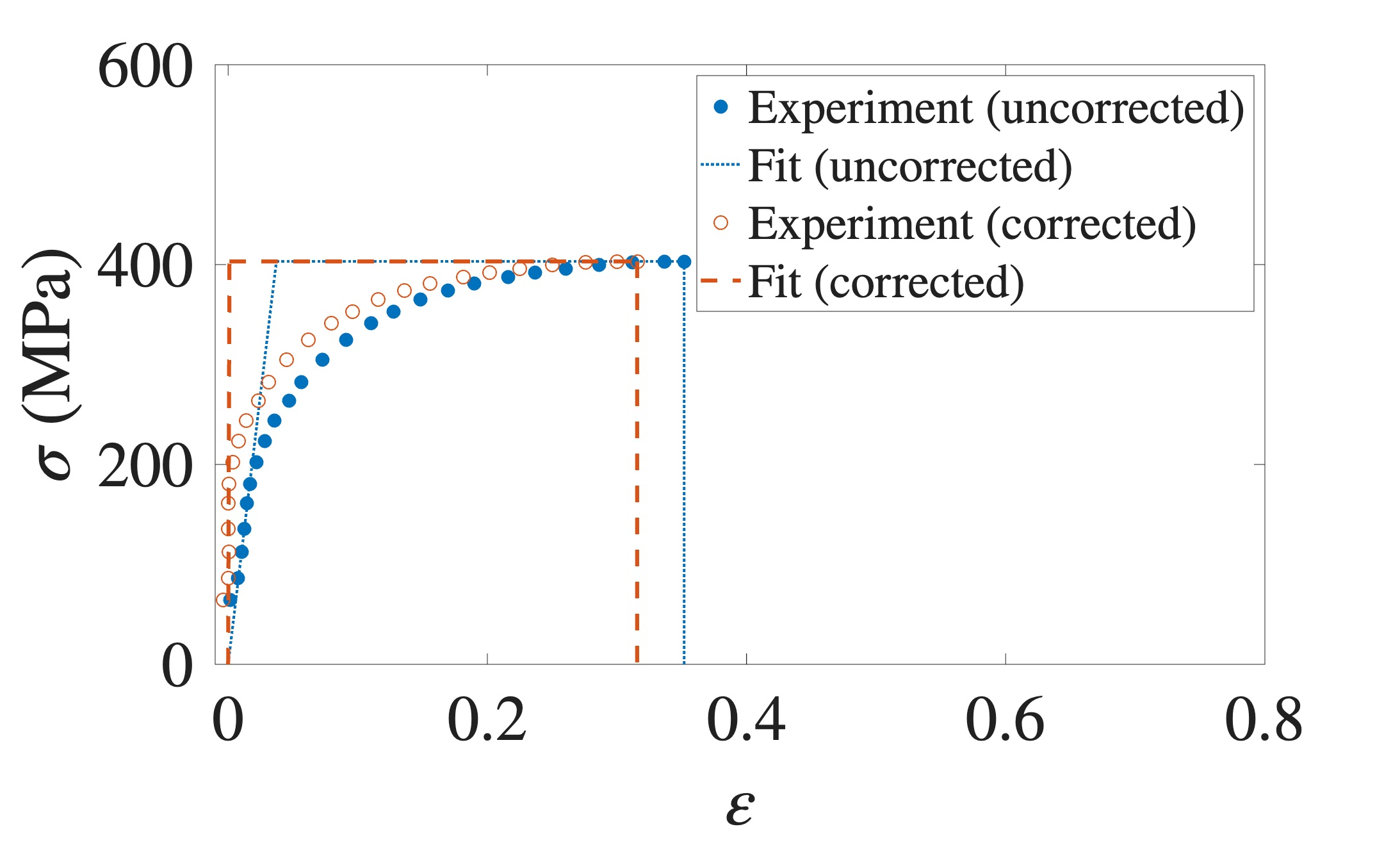}
        \subcaption{$T = 300\degC$}
        \label{fig:sub1}
    \end{subfigure}%

    \begin{subfigure}[b]{0.5\textwidth}
        \centering
        \includegraphics[width=\textwidth]{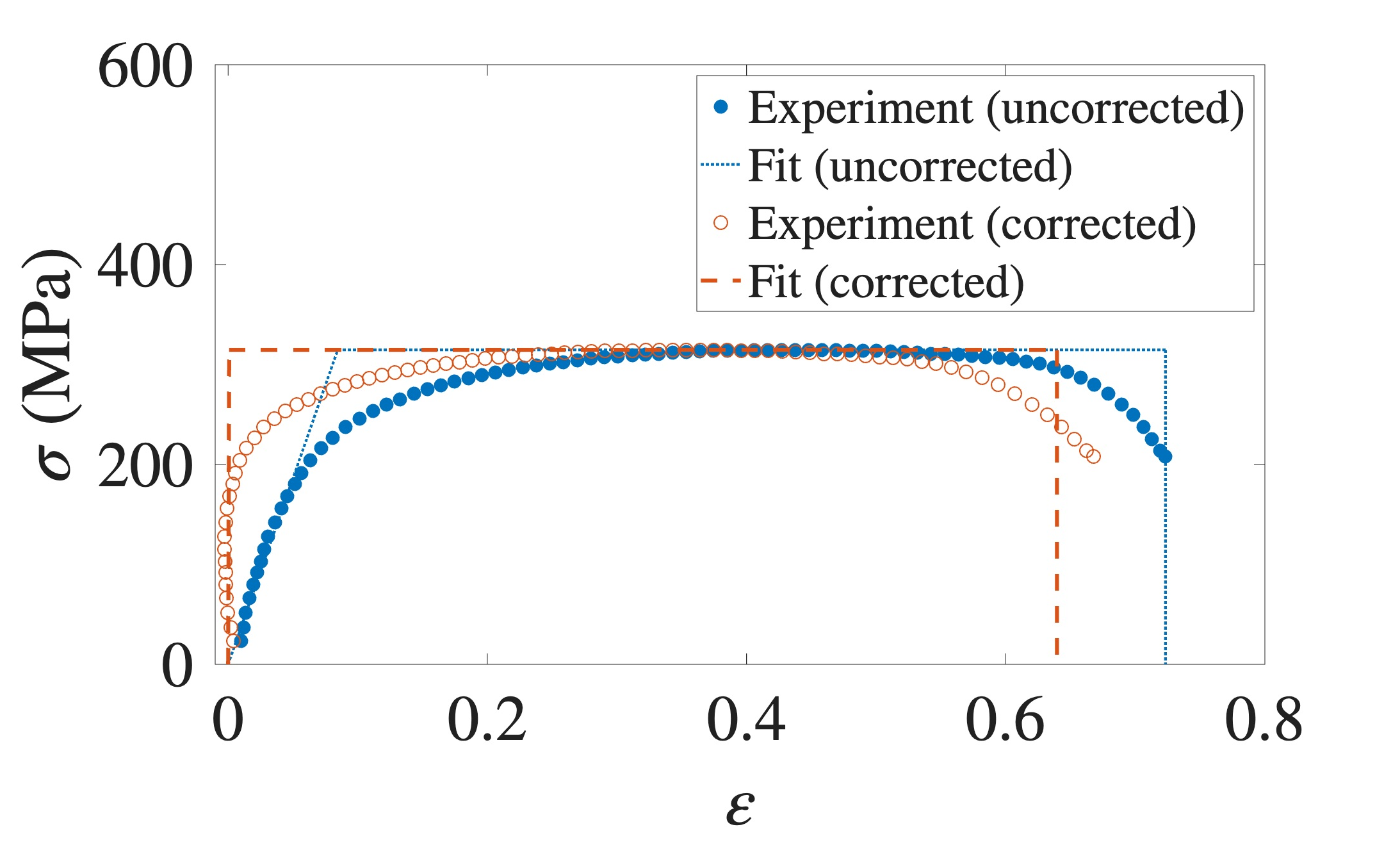}
        \subcaption{$T = 500\degC$}
        \label{fig:sub2}
    \end{subfigure}%

    \begin{subfigure}[b]{0.5\textwidth}
        \centering
        \includegraphics[width=\textwidth]{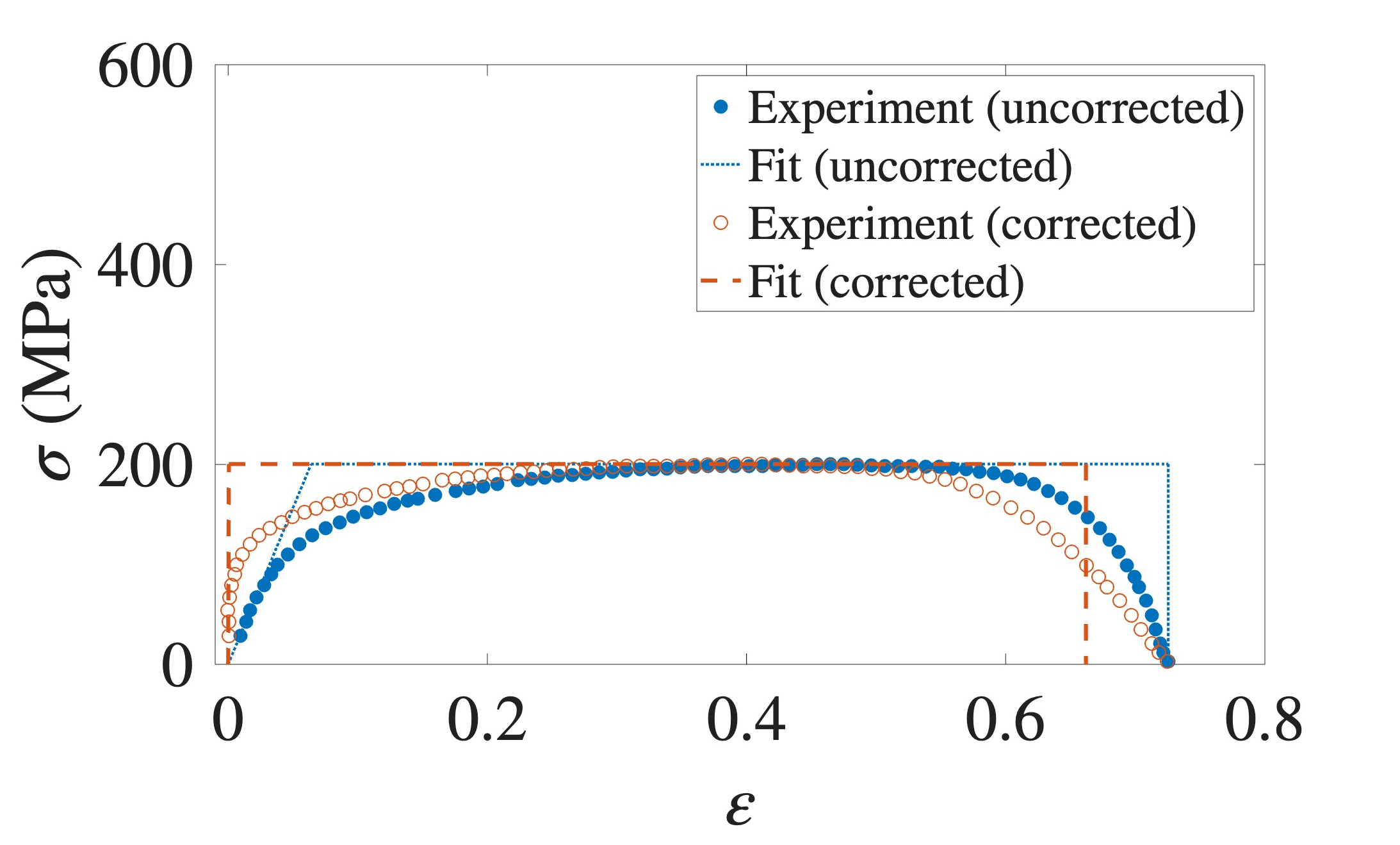}
        \subcaption{$T = 1000\degC$}
        \label{fig:sub3}
    \end{subfigure}

    \caption{Stress\PS{--}strain curves at three temperatures: $300\degC$ (top), $500\degC$ (middle), and $1000\degC$ (bottom), plotted using engineering stress. Blue filled markers indicate uncorrected digitized data points from~\cite{wirtz2016}, and blue dotted curves indicate uncorrected \ac{EPP} fits to those points. The machine-compliance correction using~\eqref{eq: machine compliance correction} is shown with orange open markers for the corrected data and orange dashed curves for the corrected fits.}
    \label{fig: stress-strain curves}
\end{figure}

\PS{
The yield strain, $\varepsilon_{y}$, and the failure strain, $\varepsilon_{\rm fail}$, are used in the \ac{PD} model in the equations for the yield stretch (see~\eqref{eq: yield stretch}) and the critical stretch (see~\eqref{eq: critical stretch}), respectively; both material properties were determined from tensile-test stress--strain curves.  
}Stress\PS{--}strain curves \PS{from} tensile tests \PS{at three temperatures---$300\degC$, $500\degC$, and $1000\degC$---}were reported in 
~\cite{wirtz2016}.
The digitized experimental data \PS{are} shown in Figure~\ref{fig: stress-strain curves}. We note that, at $300\degC$, the experimental curve does not start at the origin, whereas, at the other temperatures, data at very small strain values are missing. This may be due to backlash in the testing apparatus, as suggested in~\cite{wirtz2013}. For each of the three temperatures, we fitted an \ac{EPP} curve to the experimental data. 
\PSS{
In the EPP model, the stress, $\sigma$, increases linearly with strain, $\varepsilon$, in the elastic regime and remains constant once $\varepsilon$ reaches $\varepsilon_y$; 
in the present work, to represent failure as shown in Figure~\ref{fig: stress-strain curves}, 
$\sigma$ is further assumed to drop to zero once $\varepsilon$ reaches $\varepsilon_{\rm fail}$.} %
In addition, to account for the apparent backlash, we subsequently applied a machine-compliance correction to the curves, as described below. 

To fit an \ac{EPP} curve, we first estimated the apparent Young's modulus, $E_a$,  in the elastic regime using a linear fit constrained to pass through the origin. Due to the issue with data near the origin discussed above, the fit used data corresponding to $0.5\%$--$2.0\%$ strain. 
Next, we extracted the maximum stress, $\sigma_{\rm max}$\PS{,} and used this as an estimate for the yield stress under the \ac{EPP} assumption. From this, we obtained the yield strain as $\varepsilon_{y}= \sigma_{\rm max}/E_a$. Finally, we took the failure strain, $\varepsilon_{\rm fail}$, as the maximum strain. 
The values obtained are presented in Table~\ref{Table: Material mechanics properties} (under ``Uncorrected mechanical properties'') and the resulting \ac{EPP} curves are shown in Figure~\ref{fig: stress-strain curves}.

\begin{table}[h!]
\centering
\begin{tabular}{|c c c c c |} 
 \hline
 \multicolumn{5}{|c|}{\bf Uncorrected mechanical properties }\\
 \hline
 {Property} & \multicolumn{3}{c}{Value} & {Units} \\ 
 \cline{2-4}
 & $300\degC$ & $500\degC$ &  $1000\degC$ & \\  
 \hline 
$E_a$ & $10.8$ & $3.72$ & $3.13$ & GPa \\ 
$\sigma_{\rm max}$ & $403$ & $315$ & $200$ & MPa \\
$\varepsilon_{y}$  & $3.72\times 10^{-2}$ & $8.46\times 10^{-2}$ & $6.41\times 10^{-2}$  & --- \\
$\varepsilon_{\rm fail}$ & $0.352$ & $0.723$ & $0.725$ & --- \\ 
 \hline
 \hline
 \multicolumn{5}{|c|}{\bf Corrected mechanical properties}\\
\hline
 {Property} & \multicolumn{3}{c}{Value} & {Units} \\ 
 \cline{2-4}
 & $300\degC$ & $500\degC$ &  $1000\degC$ & \\  
 \hline 
 $E$ & $401$ & $393$ & $374$  & GPa \\
$\varepsilon_{y}$  & $1.01\times 10^{-3}$ & $8.00\times 10^{-4}$ & $5.36\times 10^{-4}$  & --- \\
$\varepsilon_{\rm fail}$  & $0.316$ & $0.640$ & $0.662$ & --- \\ 
 \hline
\end{tabular}
\caption{Mechanical properties. Uncorrected mechanical properties are obtained directly from the experimental stress\PS{--}strain data in~\cite{wirtz2016} using an \ac{EPP} fit. Corrected mechanical properties include Young's modulus, computed with~\eqref{eq: Young's modulus} (measured via impulse excitation), and the machine-compliance-corrected yield strain and failure strain.}
\label{Table: Material mechanics properties}
\end{table}

\begin{figure}[h!]
    \centering
    \includegraphics[width=0.5\textwidth, trim=0cm 0cm 0cm 0cm, clip]{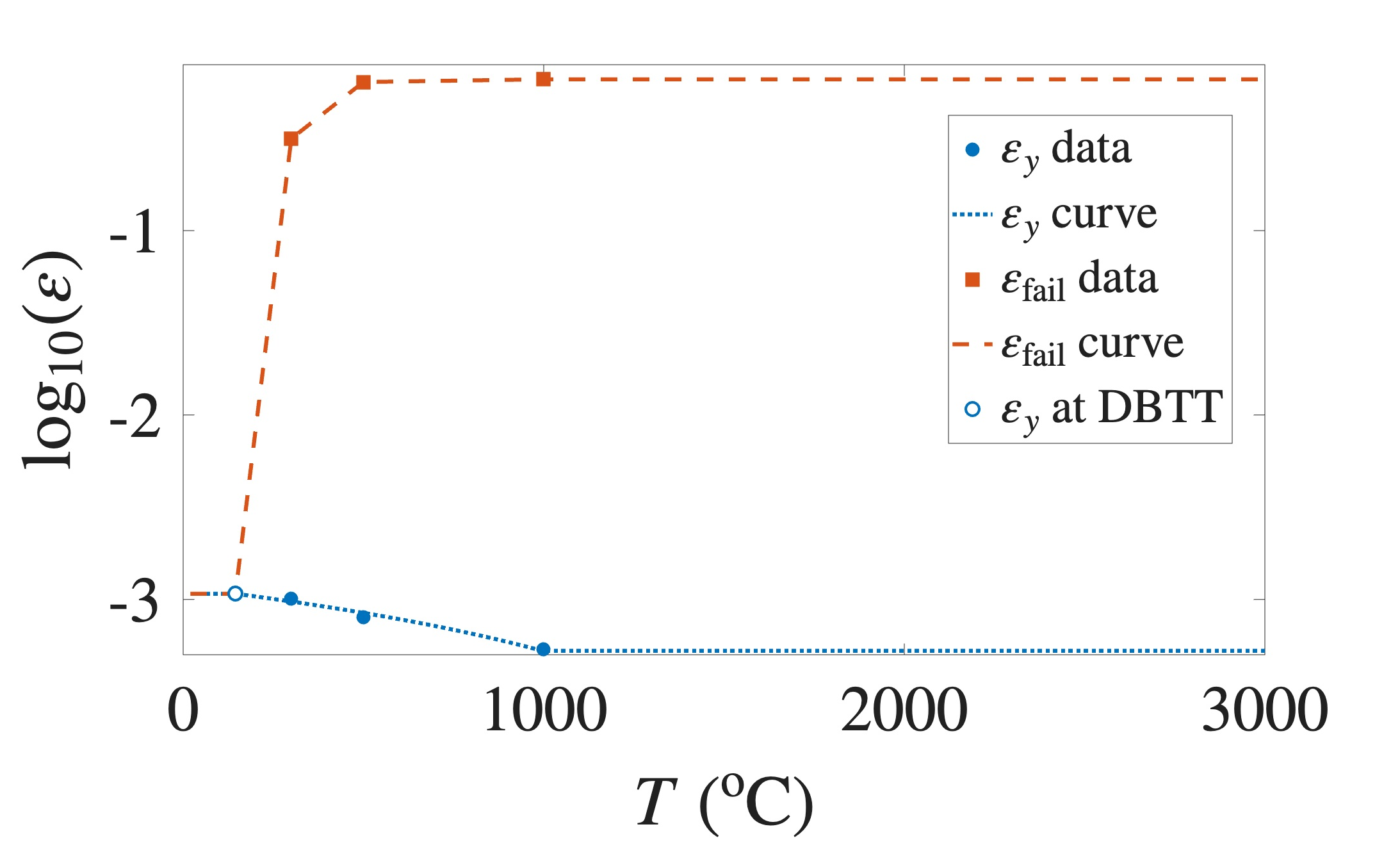}
    \caption{Temperature dependence of \PSS{the} yield strain\PSS{, $\varepsilon_y$,} and \PSS{the} failure strain\PSS{, $\varepsilon_{\rm fail}$,} shown on a semilog-$y$ plot. Filled markers indicate values obtained from \ac{EPP} fits to experimental data from~\cite{wirtz2016} after machine-compliance correction (see Table~\ref{Table: Material mechanics properties}): blue circles represent yield strain, and orange squares represent failure strain. The dotted blue line shows the linear fit to the yield-strain data and its constant extrapolation, and the open blue circle indicates the yield strain at the \ac{DBTT} predicted by this fit. The dashed orange curve shows the failure strain obtained by piecewise-linear interpolation and constant extrapolation.}
    \label{fig: Yield strain and failure strain}
\end{figure}

As observed in Table~\ref{Table: Material mechanics properties}, the apparent Young's moduli are significantly lower than the independent measurements from Section~\ref{sec: Young's modulus}. In~\cite{wirtz2013}, it was explained that specimen deformation during the tensile tests could not be measured with an extensometer due to the high temperatures and limited instrumentation; therefore, strain was inferred from  jaw-to-jaw displacement. Consequently, a machine-compliance correction is required to remove the elastic deformation of the load train from the measured engineering strain. The corrected engineering strain, $\varepsilon_{\rm corr}$, was obtained as follows: 
\begin{equation}
\label{eq: machine compliance correction}
\varepsilon_{\rm corr} = \varepsilon_{\rm meas} - \frac{\sigma}{E_a} + \frac{\sigma}{E_r},
\end{equation}
where $\varepsilon_{\rm meas}$ is the measured engineering strain,   $\sigma$ is the engineering stress, and $E_r$ is the reference (true) Young's modulus of the material, determined in this paper using~\eqref{eq: Young's modulus}. The corrected values of $\varepsilon_{y}$ and $\varepsilon_{\rm fail}$, as well as $E$ computed with~\eqref{eq: Young's modulus}, at each temperature are presented in Table~\ref{Table: Material mechanics properties} (under ``Corrected mechanical properties''), 
and the corrected \ac{EPP} curves are shown in Figure~\ref{fig: stress-strain curves}. Figure~\ref{fig: Yield strain and failure strain} shows the $\varepsilon_{y}$ and $\varepsilon_{\rm fail}$ extracted from those curves as a function of temperature. 

Because there was no room-temperature mechanical testing in~\cite{wirtz2013} or in~\cite{wirtz2016}, we used the \acf{DBTT} of $145\degC$ reported  in~\citet{yin_2021} for pure tungsten in the longitudinal orientation of a forged sample (\ac{IGP} L-T);  recrystallized samples were not tested \PS{in that study}. At this temperature, we assumed  $\varepsilon_{\rm fail}=\varepsilon_{y}$, where the value of  $\varepsilon_{y}$ at the \ac{DBTT} was obtained by linear extrapolation from the higher-temperature values. To this end, we performed a linear fit of $\varepsilon_{y}$ using the three corrected values listed in Table~\ref{Table: Material mechanics properties}, resulting in the following function: 
\begin{equation}
\label{eq: yield strain}
\varepsilon_y (T) = -6.43\times 10^{-7} \, T + 1.17 \times 10^{-3},  {\rm DBTT} \leq T\leq 1000\degC, 
\end{equation}
where the temperature, $T$, is in\PS{$\degC$}.  
Below the \ac{DBTT} and above $1000\degC$, the value of $\varepsilon_y$ was held constant because of the lack of additional data; this also prevents $\varepsilon_y$ from becoming negative at higher temperatures.  
Note that although $\varepsilon_y$ is assumed to remain constant above $1000\degC$, $E$ decreases linearly with increasing temperature (see Figure~\ref{fig: Elastic modulus}), resulting in a corresponding temperature-dependent decrease in $\sigma_{\rm max}$. 

In contrast to the linear fit used for $\varepsilon_y$, piecewise linear interpolation was adopted for $\varepsilon_{\rm fail}$, based on its value at the \ac{DBTT} and the three corrected values listed in  Table~\ref{Table: Material mechanics properties}. In the absence of further information on $\varepsilon_{\rm fail}$ at lower temperatures, the value at the \ac{DBTT} was also held constant when extrapolating down to room temperature. We note that this is a conservative choice, as $\varepsilon_{\rm fail}$ likely decreases with decreasing temperature. Similarly, $\varepsilon_{\rm fail}$ was held constant above $1000\degC$. 
We note that the \ac{DBTT} is a critical input for these simulations, evident in the substantial change in $\varepsilon_{\rm fail}$ in Figure~\ref{fig: Yield strain and failure strain}.

\subsection{Thermal properties}
\label{sec: Thermal properties}

\subsubsection{Coefficient of linear thermal expansion}

\PS{The \ac{CTE}, $\alpha$, is used in the \ac{PD} model in the equation for the thermal stretch (see~\eqref{eq: thermal stretch}).
Its temperature dependence was}
measured in~\cite{wirtz2013} using a vertical double-pushrod dilatometer. 
The reported experimental data  
cover an approximate temperature range from $300\degC$ to $900\degC$. The digitized data are shown in Figure~\ref{fig: CTE} with a linear fit to the data, resulting in the following function:
\begin{equation}
\label{eq: CTE}
\alpha(T) = 2.56\times 10^{-10} \, T + 4.74\times 10^{-6},
\end{equation}
where the temperature, $T$, is in\PS{$\degC$} and $\alpha$ is in\PS{$\degC$$^{-1}$}.  

\subsubsection{Specific heat capacity}

\PS{The specific heat capacity, $c_p$, is used in the \ac{PD} model in the governing equation for heat transfer (see~\eqref{eq: PD heat transfer}).
Its temperature dependence} 
was measured in~\cite{wirtz2013} using high-temperature differential scanning calorimetry. 
The reported experimental data cover temperatures up to $1200\,^{\circ}\mathrm{C}$.
The digitized data are shown in Figure~\ref{fig: c_p} 
\PSSS{with} 
a linear fit to the data\PSSS{,}
 resulting in the following function:
\begin{equation}
\label{eq: Specific heat capacity}
c_p(T) = 1.30\times 10^{-2} \, T + 1.42\times 10^{2},
\end{equation}
where the temperature, $T$, is in\PS{$\degC$} and $c_p$ is in J/(kg\PS{$\dotdegC$}). 

\subsubsection{Thermal conductivity}

\begin{figure}[h!]
    \centering
    \begin{subfigure}[b]{0.5\textwidth}
        \centering
        \includegraphics[width=\textwidth, trim=0cm 0cm 0cm 0cm, clip]{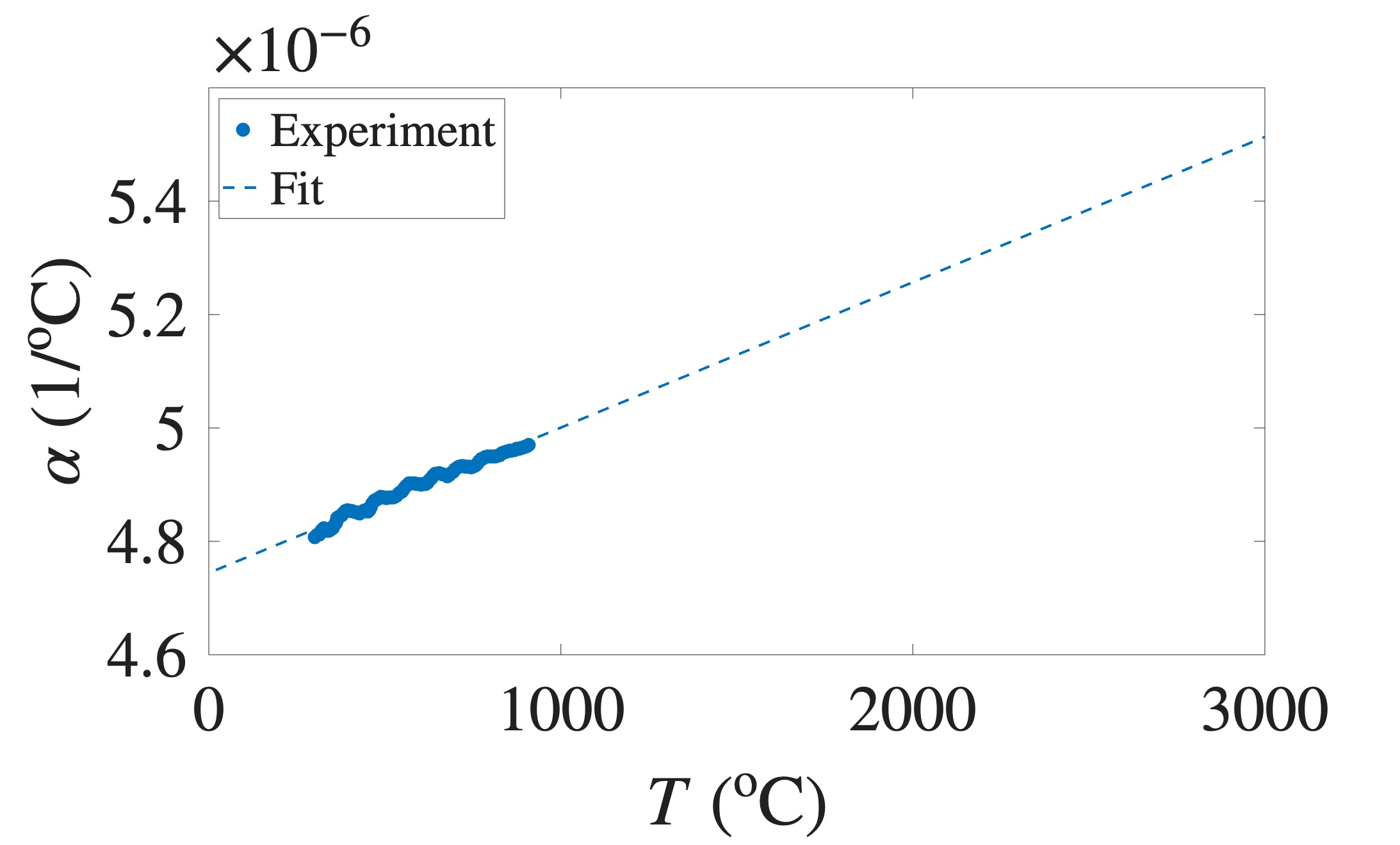}
        \subcaption{\ac{CTE}}
        \label{fig: CTE}
    \end{subfigure}%

    \begin{subfigure}[b]{0.5\textwidth}
        \centering
        \includegraphics[width=\textwidth]{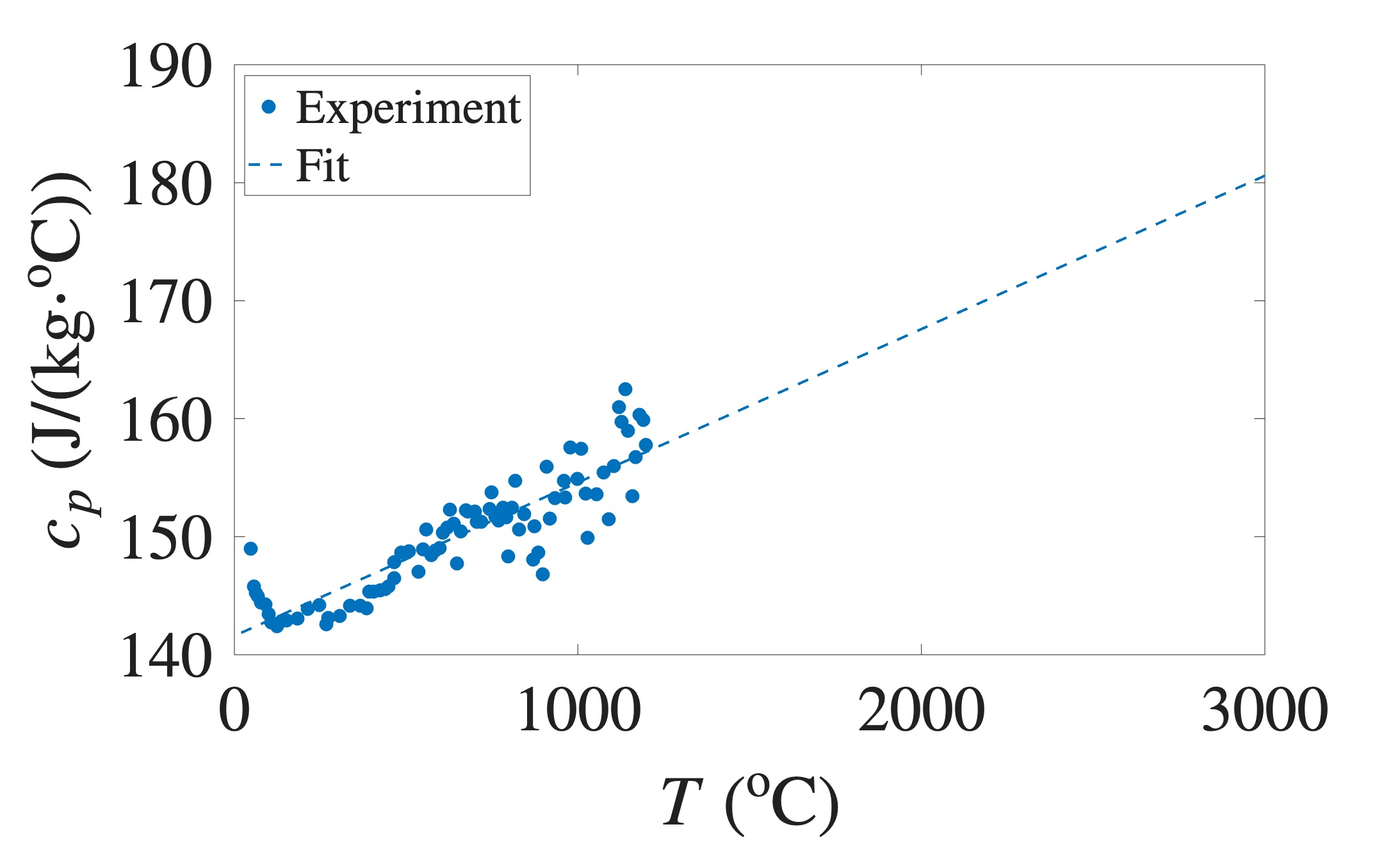}
        \subcaption{Specific heat capacity}
        \label{fig: c_p}
    \end{subfigure}%

    \begin{subfigure}[b]{0.5\textwidth}
        \centering
        \includegraphics[width=\textwidth]{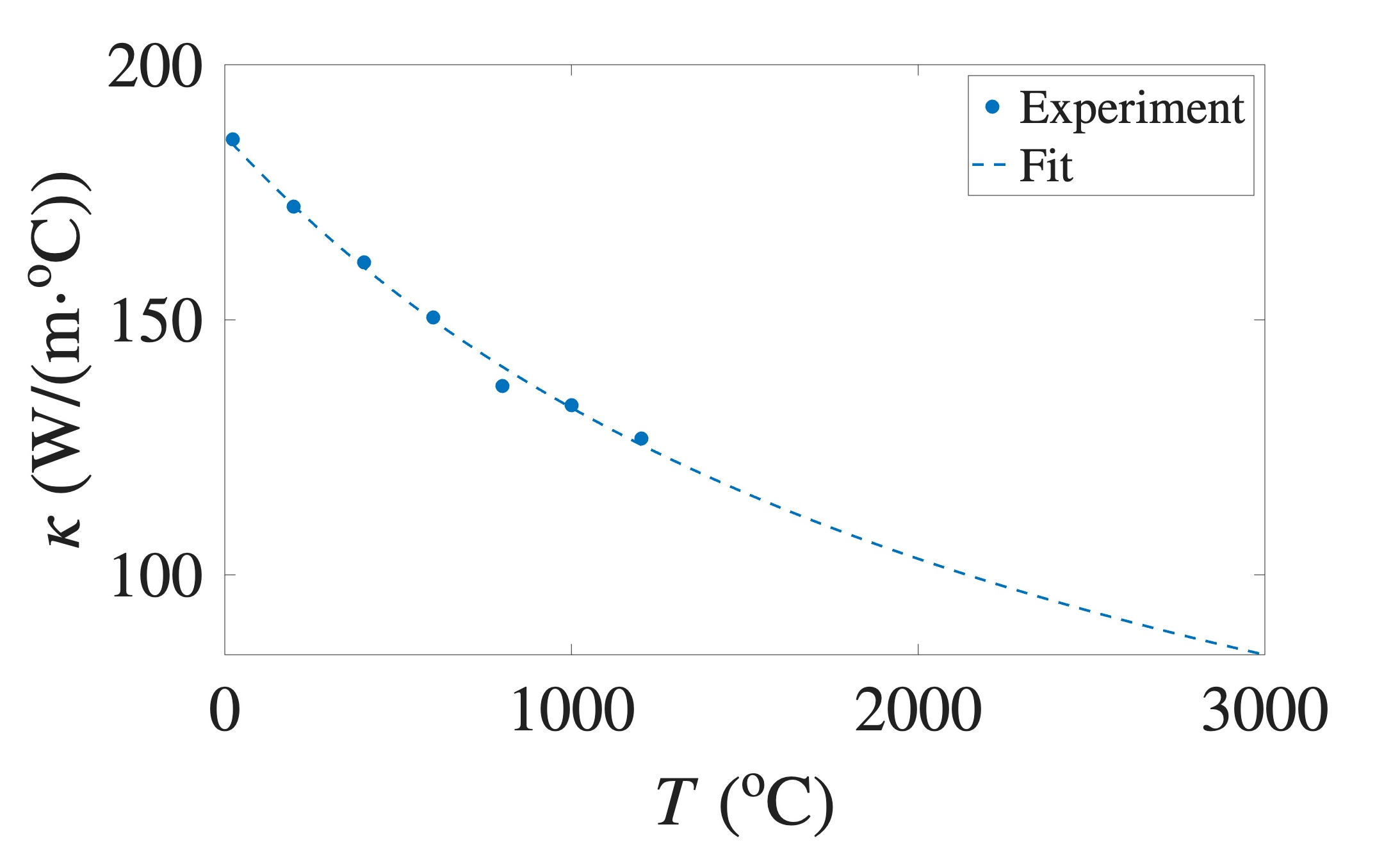}
        \subcaption{Thermal conductivity}
        \label{fig: kappa}
    \end{subfigure}

    \caption{Temperature dependence of the thermal properties. \PS{Markers indicate}  experimental data from~\cite{wirtz2013}\PS{;} dashed curves \PS{indicate} fits to the data.}
    \label{fig: thermal properties}
\end{figure}

\PS{The thermal conductivity, $\kappa$, is used in the \ac{PD} model in the equation for the thermal microconductivity  (see~\eqref{eq: micro-conductivity}).
Its temperature dependence was} 
calculated in~\cite{wirtz2013} as the product of the thermal diffusivity, the specific heat capacity, and the \PS{mass} density, where the thermal diffusivity was determined with a laser-flash device.  The reported experimental data cover temperatures up to $1200\,^{\circ}\mathrm{C}$.
In this case, a linear fit \PS{of $\kappa$ as a function of the temperature, $T$,} was deemed unsuitable%
; instead, we performed a linear fit of $1/\kappa$ as a function of $T$. Inverting this relation yields the following function:
\begin{equation}
\label{eq: Thermal conductivity}
\kappa(T) = \frac{1}{2.16\times 10^{-6} \, T + 5.37\times 10^{-3}},
\end{equation}
where $T$ is in\PS{$\degC$} and \PS{$\kappa$} is in W/(m\PS{$\dotdegC$}). The digitized data and nonlinear fit are shown in Figure~\ref{fig: kappa}.

\subsection{Physical properties}
\label{sec: Physical properties}

\begin{figure}[h!]
    \centering
    \includegraphics[width=0.5\textwidth, trim=0cm 0cm 0cm 0cm, clip]{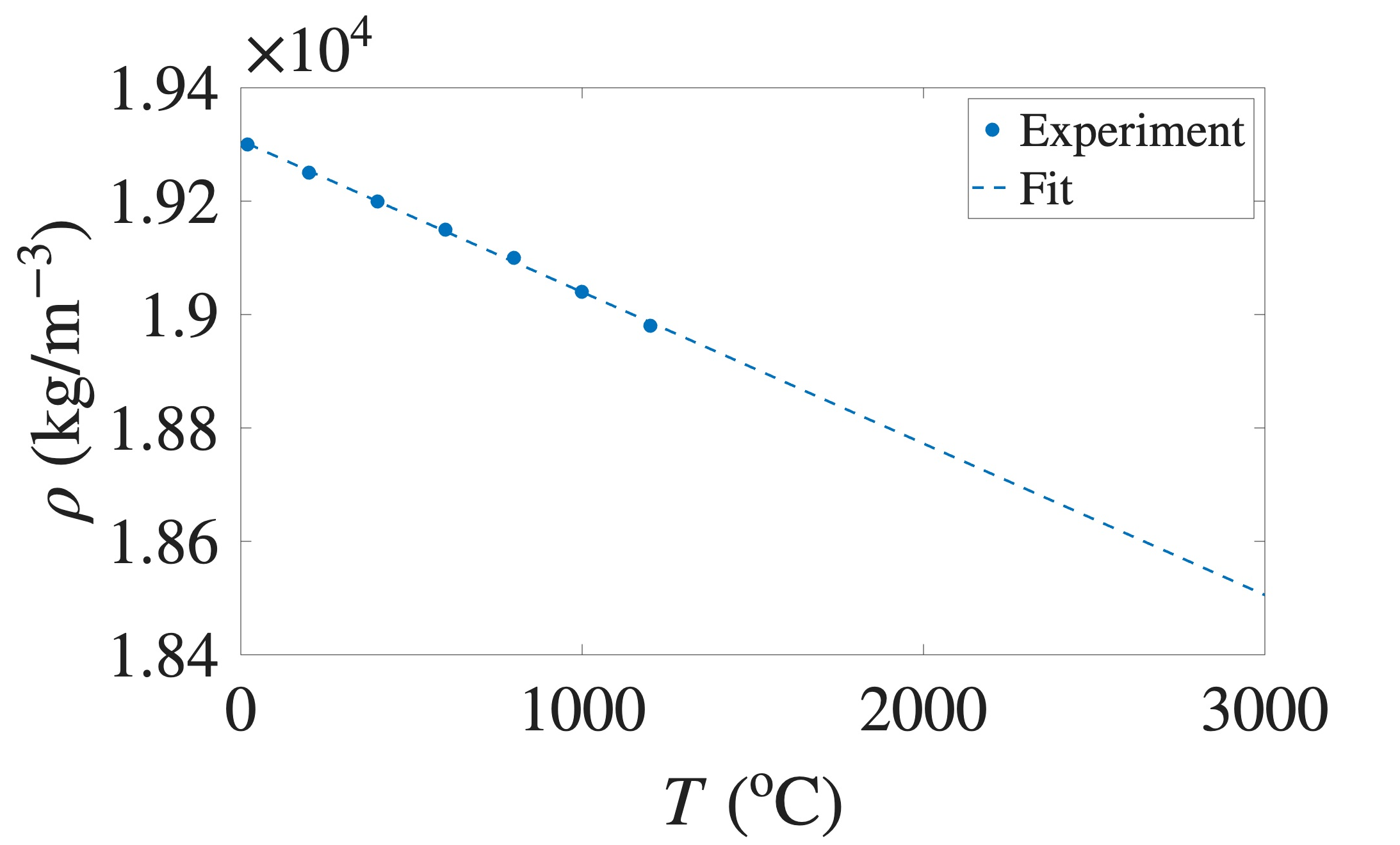}
    \caption{Temperature dependence of the \PS{mass} density. Markers indicate reported experimental data in~\cite{wirtz2013}; the dashed line indicates the linear fit.}
    \label{fig: Density}
\end{figure}

\PS{The mass density, $\rho$, is used in the \ac{PD} model in the governing equations for mechanics (see~\eqref{eq: PD mechanics}) and heat transfer (see~\eqref{eq: PD heat transfer}).
Its temperature dependence} 
was reported in~\cite{wirtz2013} based on literature values. The reported values are shown in Figure~\ref{fig: Density} with a linear fit to the data, resulting in the following function:
\begin{equation}
\label{eq: density}
\rho(T) = -2.67\times 10^{-1} \, T + 1.93\times 10^{4},
\end{equation}
where the temperature, $T$, is in\PS{$\degC$} and $\rho$ is in kg/\PS{m$^{3}$}. 

\section{E-beam simulation settings}
\label{sec:sim}

\begin{figure}[h!]
    \centering
    \includegraphics[width=0.45\textwidth, trim=0cm 0cm 0cm 1.2cm, clip]{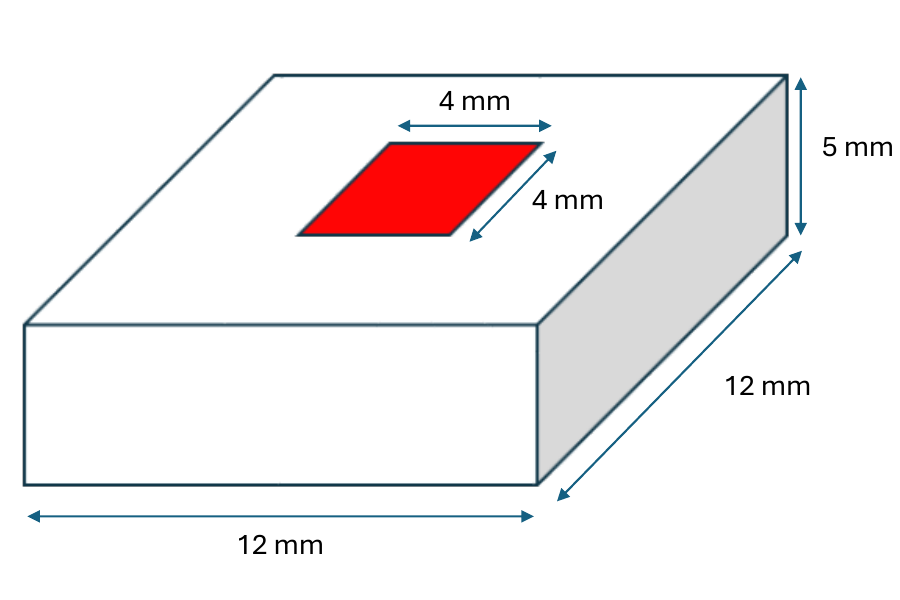}
    \caption{
    Schematic of the tungsten block subjected to heat flux~\citep{wirtz2013}. The red region on the top surface indicates the area exposed to the applied heat flux, imposed in the simulation through a temperature boundary condition.}
    \label{figS4: Wirtz case geometry}
\end{figure}

Following the experimental e-beam test conditions from \cite{wirtz2013} and~\cite{wirtz2016}, we considered a tungsten block with dimensions of $12$ $\times$ $12$ $\times$ $5$~mm, subjected to a heat flux at the center of its top surface over a $4$ $\times$ $4$~mm square region (see Figure~\ref{figS4: Wirtz case geometry}). 
The e-beam raster pattern %
was not included in the simulations because it was explicitly designed to produce  spatially homogeneous time-averaged \PSSS{thermal} loading across the square region. 
The base temperature was varied between 
23$\degC$ and $1200$\PS{$\degC$}
and the heat flux was varied between $0.20$~GW/m$^2$ and $1.80$~GW/m$^2$.

\begin{figure}[h!]
     \centering
         \centering
    \includegraphics[width=0.48\textwidth, trim=0cm 0cm 0cm 0cm, clip]
    {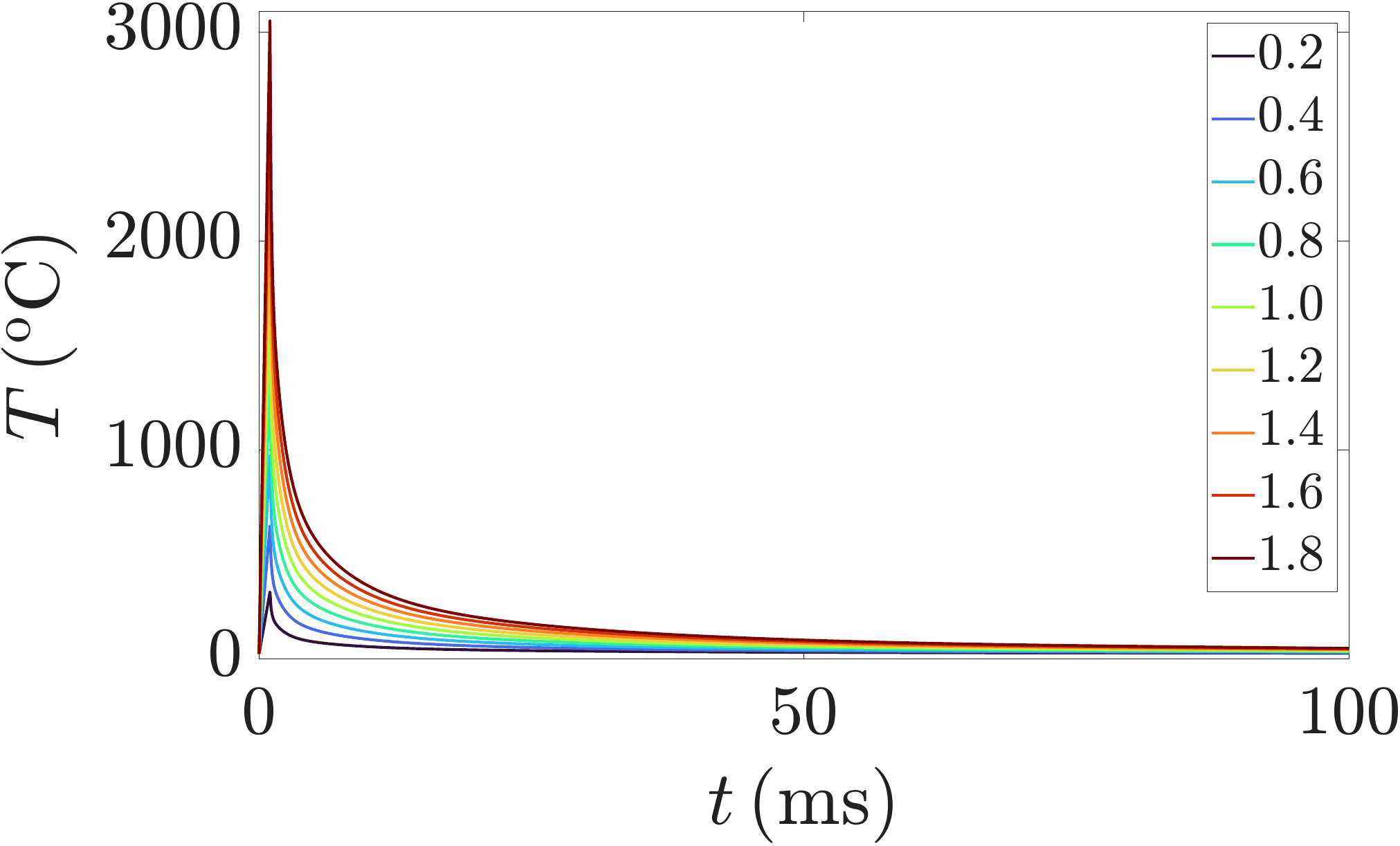}\\
    \caption{Interpolated temperature profiles  for a single heat pulse, obtained from ANSYS-generated data at a fixed base temperature $T_0 = 23$\PS{$\degC$} for different heat flux 
    \PS{values (given in the legend in GW/m$^2$).} 
    These profiles were used as Dirichlet boundary conditions in the CabanaPD simulations.}
       \label{fig: T interpolation}
\end{figure}

\PS{Heat flux} boundary conditions are not currently available in CabanaPD; therefore, transient thermal simulations
were run in the commercial finite element analysis tool ANSYS Mechanical (version 2023~R2) to convert the prescribed heat flux (Neumann) boundary condition into \JW{a surface-averaged and} time-dependent temperature (Dirichlet) boundary condition for use in CabanaPD. Specifically, transient thermal simulations of an octant of the system were conducted \PS{at room temperature} over 
$100$ ms to obtain the time-dependent temperature profiles \JW{of the average temperature of the 16~mm$^2$ heated region.}
The heat flux was applied as a constant value %
for 
$t=0{-}1$ ms and then set to $0$ W/m${^2}$ for $t=1{-}100$ ms to match the experimental thermal cycling reported in~\cite{wirtz2013}. The ANSYS-generated data points were then fit %
to obtain a piecewise-continuous interpolation function for the temperature as a function of time for a single heat pulse. This function was subsequently imposed as a Dirichlet boundary condition in the  CabanaPD simulations over the above-described square region extending a depth of $\delta$ below the top surface, where $\delta$ is the \ac{PD} horizon. 
The resulting profiles are shown in Figure~\ref{fig: T interpolation}. 
We note that, for cases with higher base temperatures, the temperature profile computed at room temperature was shifted \PSS{by the difference between the base temperature and room temperature,} 
rather than recomputed. 
 
The CabanaPD simulations used the meshfree discretization approach described in Section~\ref{sec: Meshfree approach}, \SAM{with particles} initialized \SAM{on} structured computational grids
using \SAM{uniform} spacings of $\Delta x$ = 0.24, 0.18, 0.145, 0.12, and 0.10 mm \SAM{for the convergence shown in the next section}.
These spacings were chosen using a fixed horizon of \PS{$\delta = 0.72$~mm,} 
taken from the coarsest case with $m = \delta / \Delta x= 3$, and held fixed under \PS{grid} refinement for each integer $m$ up to $m = 7$. 
This resulted in systems with a total of $50{,}000$; $125{,}692$; $241{,}115$; $420{,}000$, and \PS{$645{,}888$} particles, respectively.
Each particle position was randomly perturbed by up to $\pm \Delta x/2$ in each coordinate direction to break numerical symmetry and thereby promote damage localization. 
A zero-displacement constraint was applied to the bottom surface
in addition to the thermal boundary condition prescribed on the top surface. All simulations were run on the OLCF Frontier machine using AMD MI250X GPUs. 
\PS{The primary output of interest is the damage computed with~\eqref{eq: total damage discrete}, restricted to particles within a $9$ $\times$ $9$~mm square region at the center of the top surface and away from the  lateral boundaries.}  
Restricting the \PS{computation} of damage to the top surface \PS{and away from the lateral boundaries} 
helps both \PS{to} align with the experimental \PS{surface} measurements \PS{(obtained using  optical microscopy)}   
\PS{and to mitigate boundary-related artifacts in the computed damage}  
arising from the lack of surface\PS{-}effect correction in CabanaPD. %

For time integration of the mechanical equation of motion, \PS{a} time-stepping scheme \PS{alternating} between  quasi-static \PS{(\ac{ADR})} and explicit \PS{(velocity-Verlet)} updates was adopted. For each \ac{ADR} time interval, \ac{ADR} subcycles were run until the relative change in displacement fell below a convergence tolerance of $1\times 10^{-10}$\PSS{, without}
allowing bond breaking, followed by 100 velocity-Verlet steps \PS{to allow} potential damage propagation. 
\PSS{A stable explicit mechanical time step is calculated using~\eqref{eq: time step mechanics} with a safety factor of $\eta_{\rm m} = 0.85$. This time step depends on the grid spacing; for the coarsest system ($m = 3$), for example, $\Delta t_{\rm m} = 0.041~\mu$s.}
Heat transfer was \PS{advanced using} explicit forward Euler time-stepping\PS{,} with a step size matching the \ac{ADR} \PS{increment} or\PS{, when necessary, with sub-stepping using a stable thermal time step.} 
\PSS{A stable thermal time step is calculated using~\eqref{eq: time step heat trasnfer} with a safety factor of $\eta_{\rm th} = 0.85$, analogous to the mechanical time step. This time step also depends on the grid spacing; for the coarsest system ($m = 3$), for example, $\Delta t_{\rm th} = 0.5$ ms,} 
notably four orders of magnitude larger than the mechanical time step. 
Heat transfer was not advanced during the explicit \PSS{mechanical steps} due to this large disparity in \PSS{time-step} size.
The \ac{ADR} steps were limited to $4000 \Delta t_{\rm m}$ during the initial $1.5$ ms  
\PSS{of each heat pulse, corresponding to the portion of the cycle with non-zero heat flux, plus an additional 0.5 ms,} 
in order to efficiently converge the 
\PSS{mechanical response; this limit was determined by trial and error.} 
The \ac{ADR} step size was then increased exponentially through the rest of each cycle and reset at the start of the subsequent cycle.

\section{Results and Discussion}
\label{sec::results}

Using the experimental material properties described in Section~\ref{sec::param} and the simulation settings described  in Section~\ref{sec:sim},  
CabanaPD was used to predict cracking and damage 
in recrystallized tungsten under cyclic e-beam thermal loading. 

\begin{figure}[h!]
     \centering
    \includegraphics[width=0.5\textwidth]{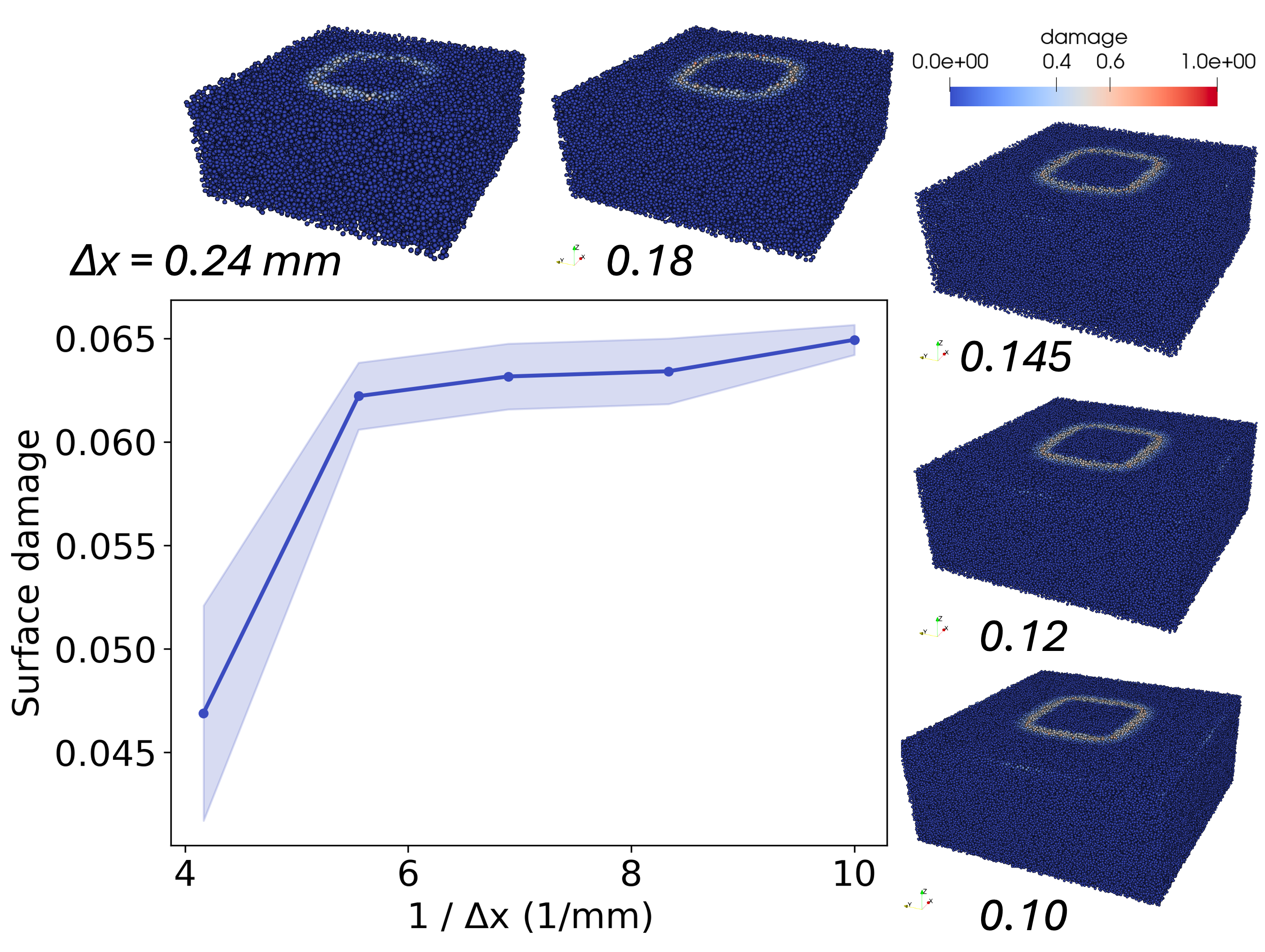}
    \caption{CabanaPD surface damage as a function of system size for $T_0 = 23$\PSS{$\degC$} and a heat flux of $ 1.20$~GW/m$^2$ after one cycle. Markers indicate the average \PSS{over} ten separate realizations at each resolution\PSS{, and the} filled region indicates one standard deviation. Insets show representative systems at each resolution\PS{,} colored by 
    \PS{the particle} damage \PS{ field (see~\eqref{eq: damage discrete})} 
    and labeled by $\Delta x$.
    }
    \label{fig:converge}
\end{figure}

To determine an adequate particle resolution for reliable damage prediction, the convergence of the surface damage profile after one cycle was first evaluated by varying the particle resolution. This quantity of interest was chosen for the convergence study for simplicity and reduced computational cost, as compared to the ultimate predictions of damage across multiple cycles. Figure~\ref{fig:converge} shows this convergence for a base temperature of $23\degC$ and a heat flux of $1.2$ GW/m$^2$, using the grid spacings detailed in Section~\ref{sec:sim}. 
The mechanical and thermal time step sizes were adjusted consistently with the grid resolution according to~\eqref{eq: time step mechanics} and~\eqref{eq: time step heat trasnfer}, respectively, and were then used in the alternating scheme described in the previous section.
Ten separate realizations, each with a unique local randomization of the particle positions (with a different random seed), were run for each system size. Based on these results, $\Delta x = 0.145$~mm was considered sufficiently converged; therefore, this grid spacing was selected for the \PSSS{remaining} %
simulations. 

With the particle resolution established, CabanaPD simulations up to 100 cycles were run across all considered heat fluxes and base temperatures. The results were analyzed and classified into three primary categories, with representative cases shown in Figure~\ref{fig:classify}: no cracks, surface cracks, and through cracks. 
\PSSS{C}ases with no cracks showed nearly zero damage (see Figure~\ref{fig:classify(a)}) or only small amounts of particle damage (see Figure~\ref{fig:classify(b)}). 
\PSSS{At} \SAM{the opposite extreme, through cracks were} 
\PSSS{identified}
\SAM{as those extending all the way to the base of the sample (see Figure}~\ref{fig:classify(g)}).
Surface cracks all showed connected regions of particle damage above 0.50, including many distinct crack morphologies, but \PSSS{these remained}
\SAM{well} within the top half of the sample in all cases, \SAM{in contrast to through cracks}. In some cases, a crack grew across the entire region where the e-beam pulse was applied (see Figure~\ref{fig:classify(e)}), while in \SAM{other} cases it did not.

Given these categories, we summarize the CabanaPD simulation results in \PSSS{the}  upper panel of Figure~\ref{fig:damage_map}.
All results are taken at $t = 10~$s.
The shaded gray region denotes combinations of base temperature and heat flux for which the peak temperature would exceed the melting point of tungsten ($3422\degC$);
because melting is not considered in CabanaPD, these conditions were intentionally excluded from the simulations.

\newcommand{\subfigwidth}{0.48\linewidth}
\newcommand{\imgwidth}{1.0\linewidth}
\begin{figure}[h!]
    \centering
     {\bf No cracks}\\[0.1in]
    \begin{subfigure}[t]{\subfigwidth}
        \centering
        \includegraphics[width=\imgwidth, trim=0cm 0cm 0cm 0cm,clip]{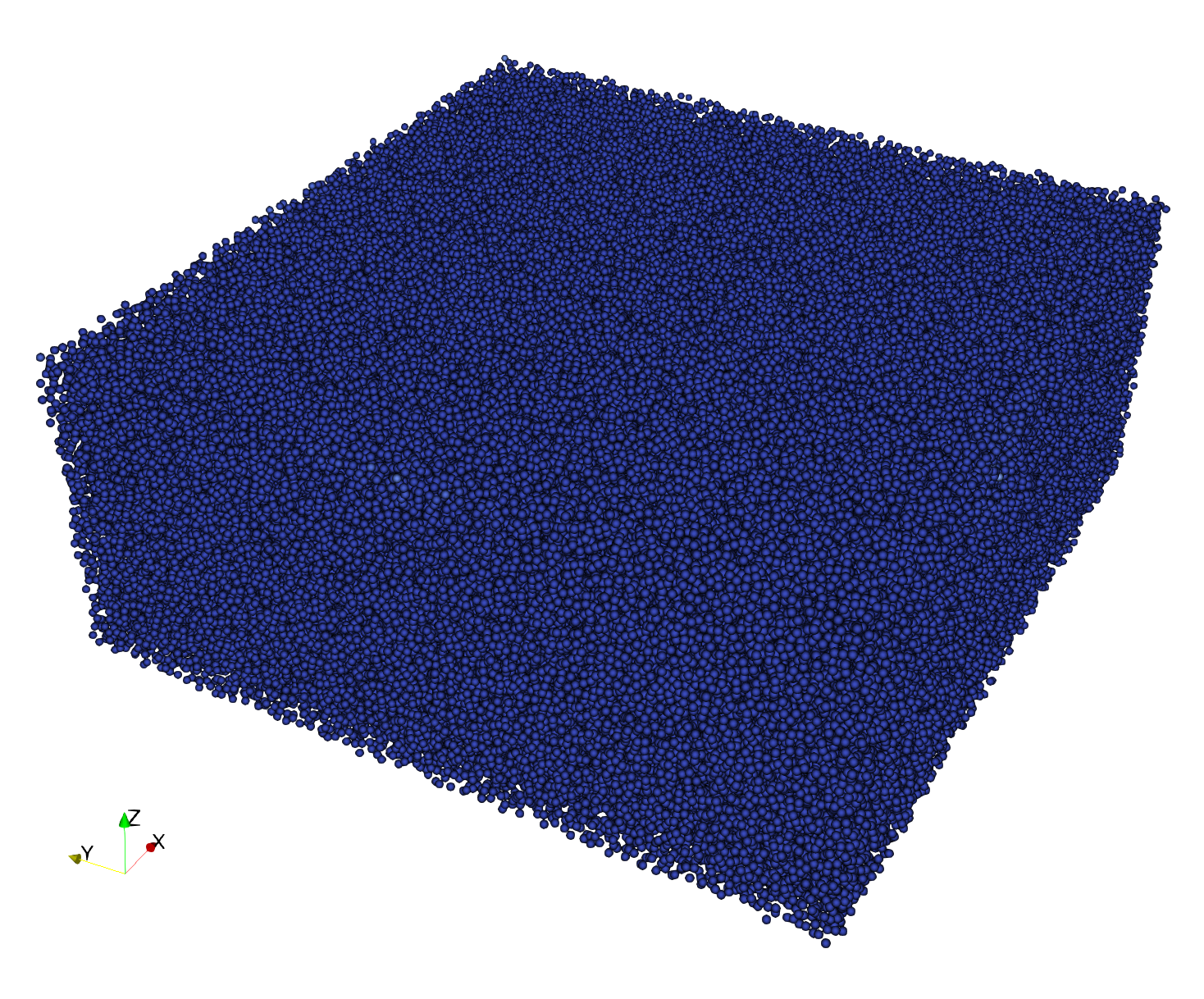}
        \caption{$0.2$ GW/m$^2$, 23$\degC$}
        \label{fig:classify(a)}
    \end{subfigure}
    \begin{subfigure}[t]{\subfigwidth}
        \centering
        \includegraphics[width=\imgwidth, trim=0cm 0cm 0cm 0cm,clip]{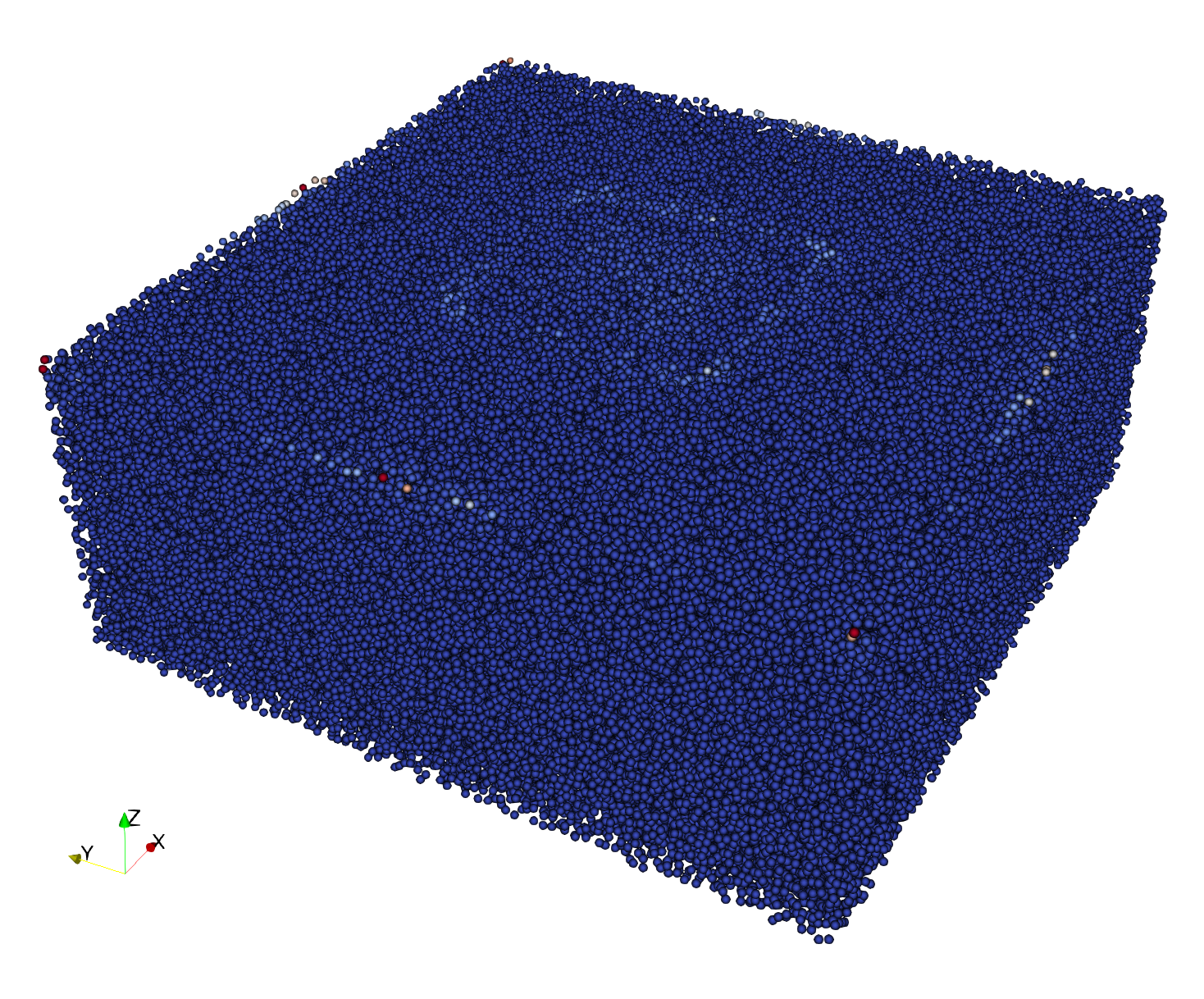}
        \caption{$0.4$ GW/m$^2$, 23$\degC$}
        \label{fig:classify(b)}
    \end{subfigure}\\[0.05in]
\noindent\rule{\linewidth}{2pt}
     {\bf Surface cracks}\\[0.1in]
    \begin{subfigure}[t]{\subfigwidth}
        \centering
        \includegraphics[width=\imgwidth, trim=0cm 0cm 0cm 0cm,clip]{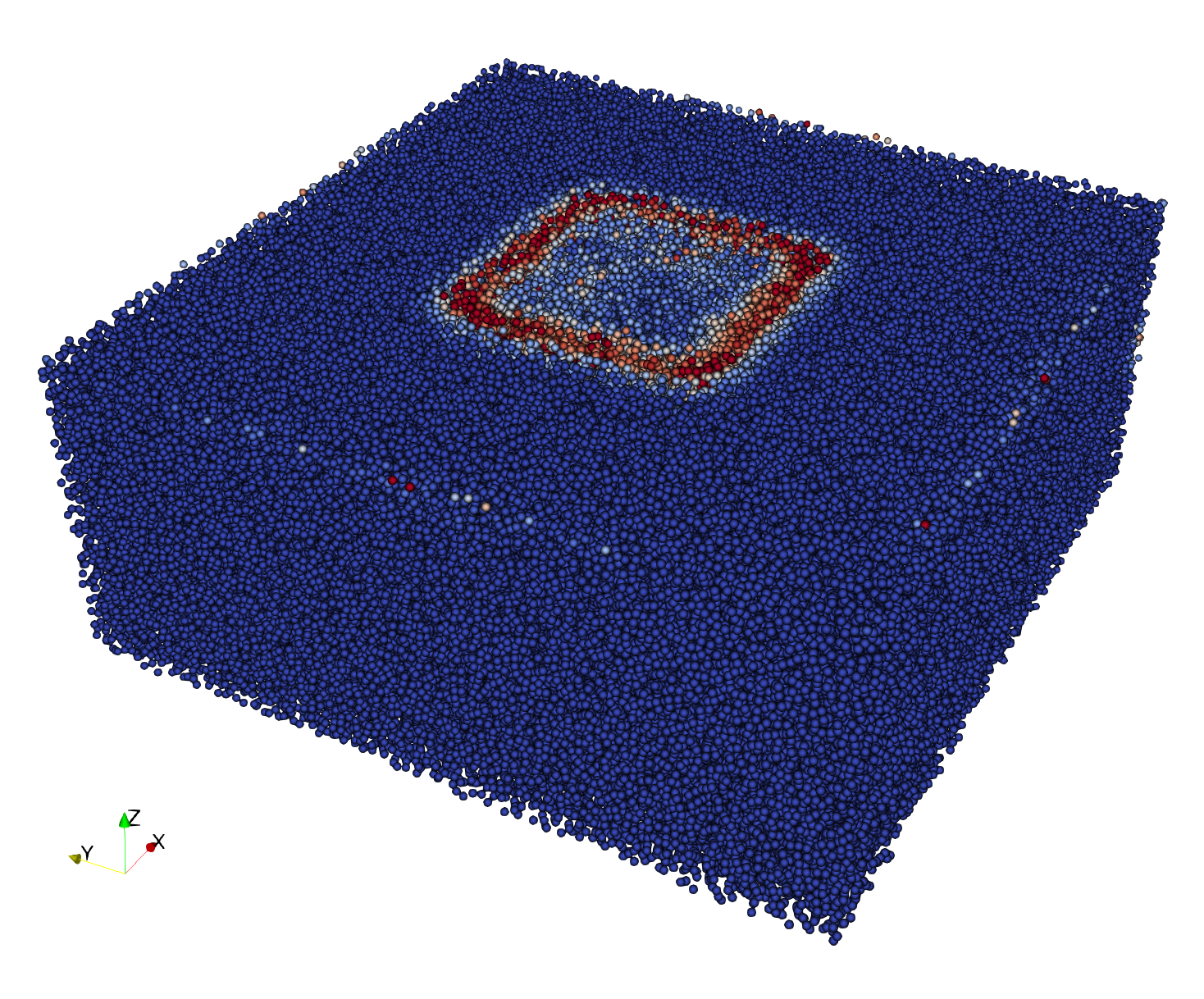}
        \caption{$0.6$ GW/m$^2$, 23$\degC$}
        \label{fig:classify(c)}
    \end{subfigure}
    \begin{subfigure}[t]{\subfigwidth}
        \centering
        \includegraphics[width=\imgwidth, trim=0cm 0cm 0cm 0cm,clip]{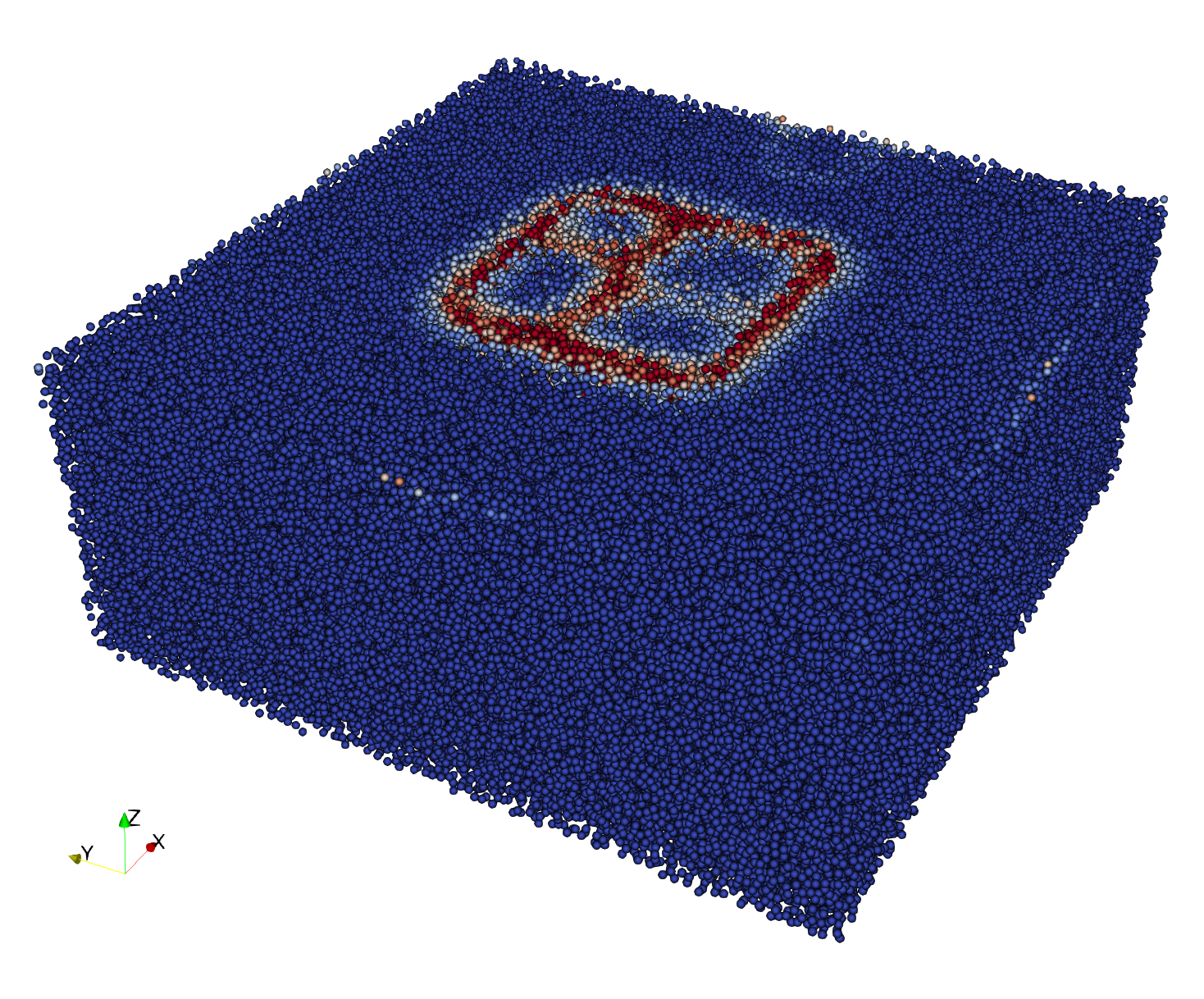}
        \caption{$0.8$ GW/m$^2$, 100$\degC$}
        \label{fig:classify(d)}
    \vspace{2ex}
    \end{subfigure} 
    \begin{subfigure}[t]{\subfigwidth}
        \centering
        \includegraphics[width=\imgwidth, trim=0cm 0cm 0cm 0cm,clip]{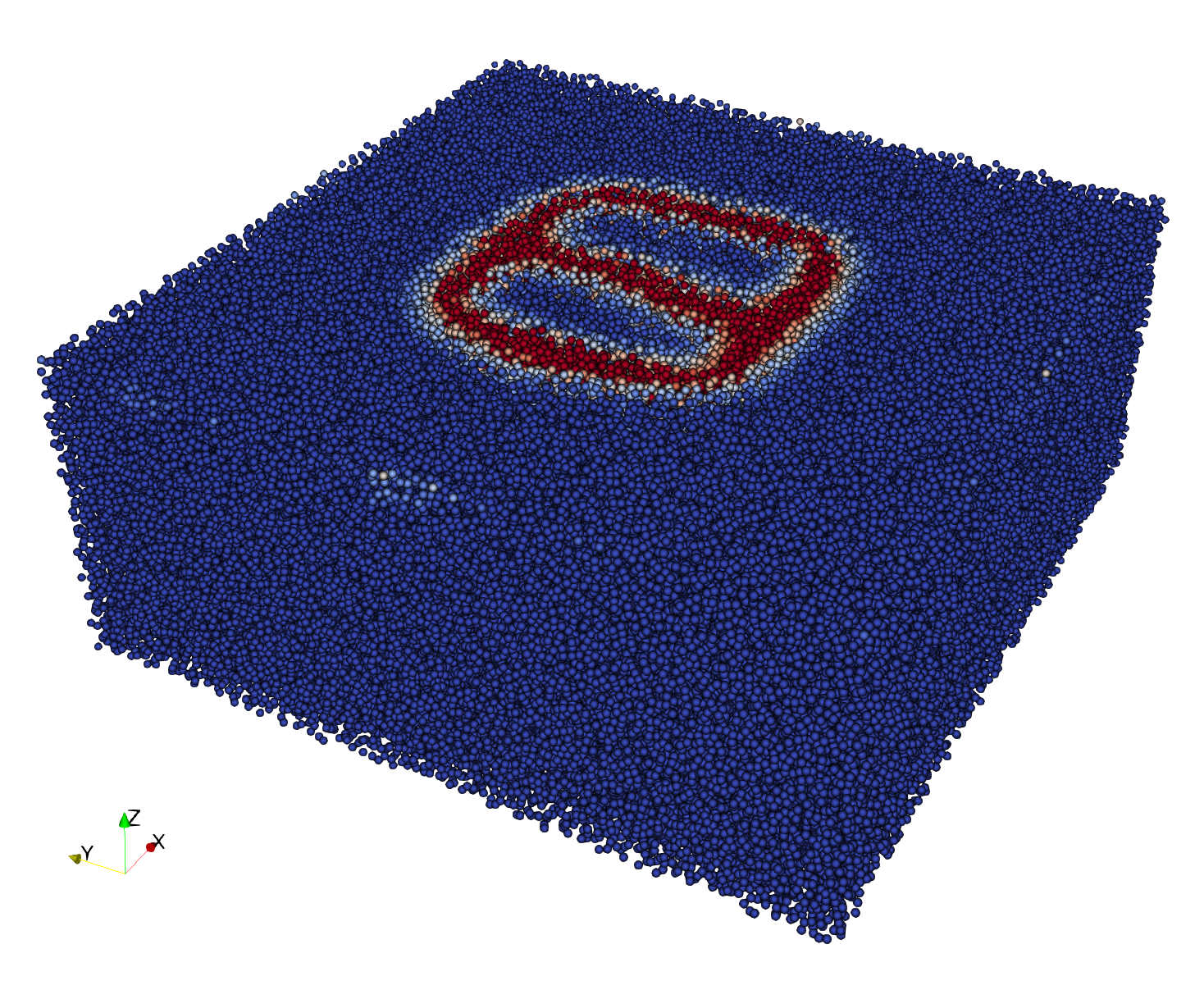}
        \caption{$1.2$ GW/m$^2$, 23$\degC$}
        \label{fig:classify(e)}
    \end{subfigure}
    \begin{subfigure}[t]{\subfigwidth}
        \centering
        \includegraphics[width=\imgwidth, trim=0cm 0cm 0cm 0cm,clip]{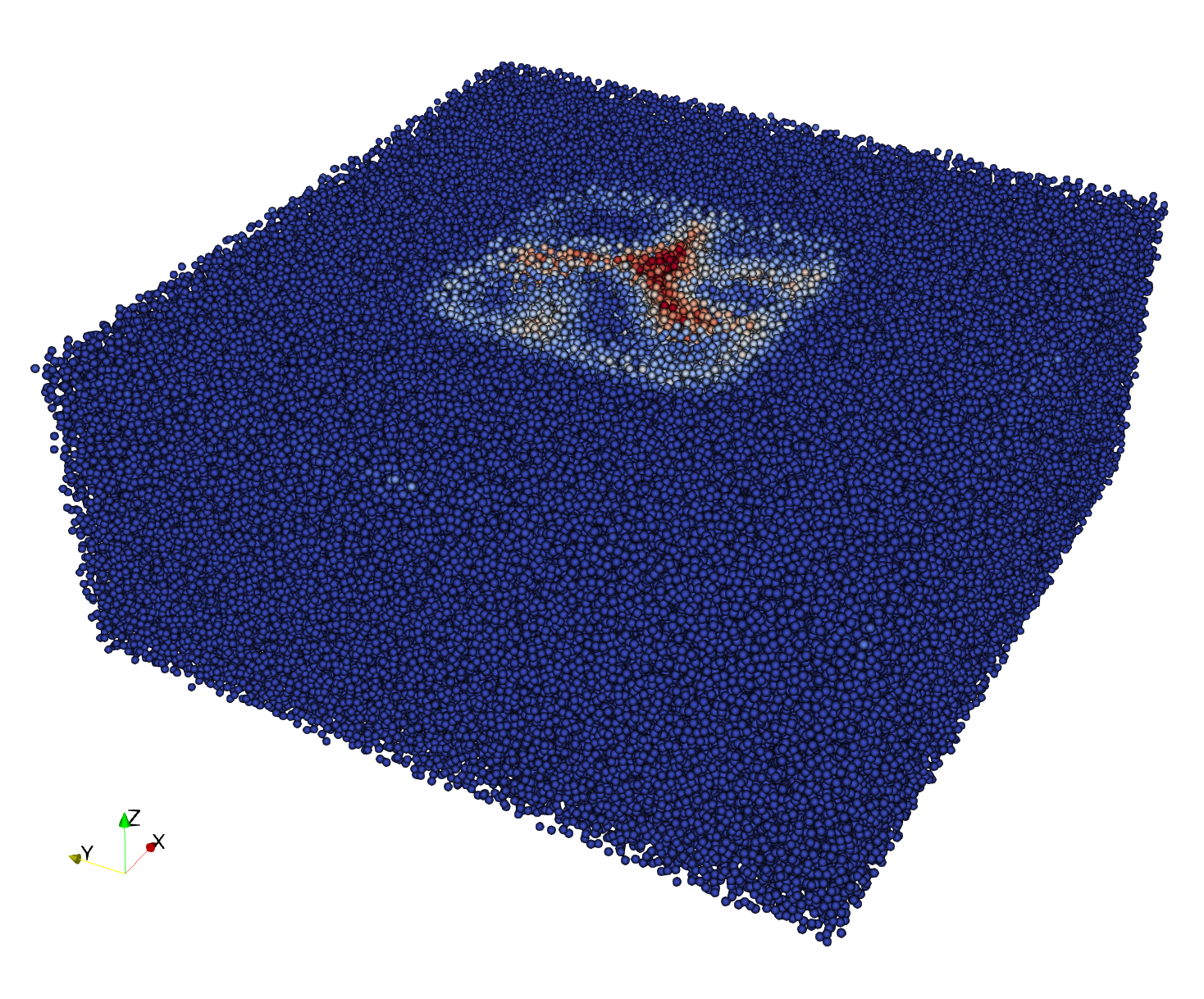}
        \caption{$1.6$ GW/m$^2$, 100$\degC$}
        \label{fig:classify(f)}
    \end{subfigure}\\[0.05in]
\noindent\rule{\linewidth}{2pt}
    \centering
     {\bf Through cracks}\\[0.1in]
    \begin{subfigure}[t]{\subfigwidth}
        \centering
        \includegraphics[width=\imgwidth, trim=0cm 0cm 0cm 0cm,clip]{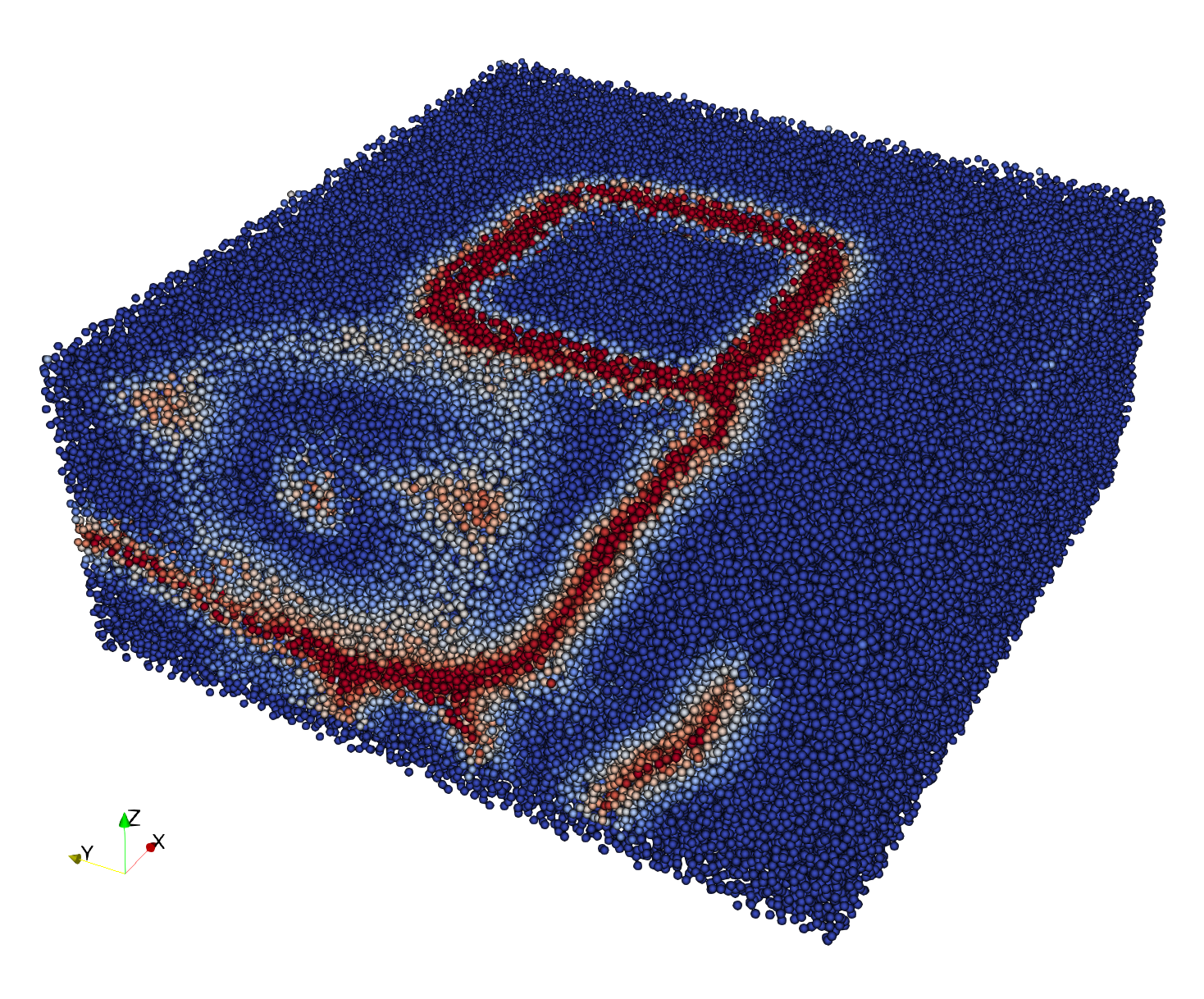}
        \caption{$1.8$ GW/m$^2$, 23$\degC$}
        \label{fig:classify(g)}
    \end{subfigure}
    \begin{subfigure}[t]{\subfigwidth}
        \centering
        \includegraphics[width=\imgwidth, trim=0cm 0cm 0cm 0cm,clip]{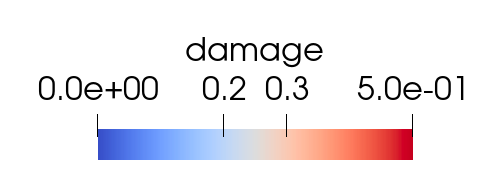}
    \end{subfigure}\\[0.05in]
\noindent\rule{\linewidth}{2pt}
    \caption{\PSS{Representative s}imulation results after $100$ cycles
    for different heat flux and base temperature values (given in each subfigure caption), organized by crack classification: no cracks, surface cracks, and through cracks. 
    All subfigures share a common colorbar for damage.}
    \label{fig:classify}
\end{figure}

\begin{figure}[h!]
     \centering
    \includegraphics[width=0.5\textwidth]{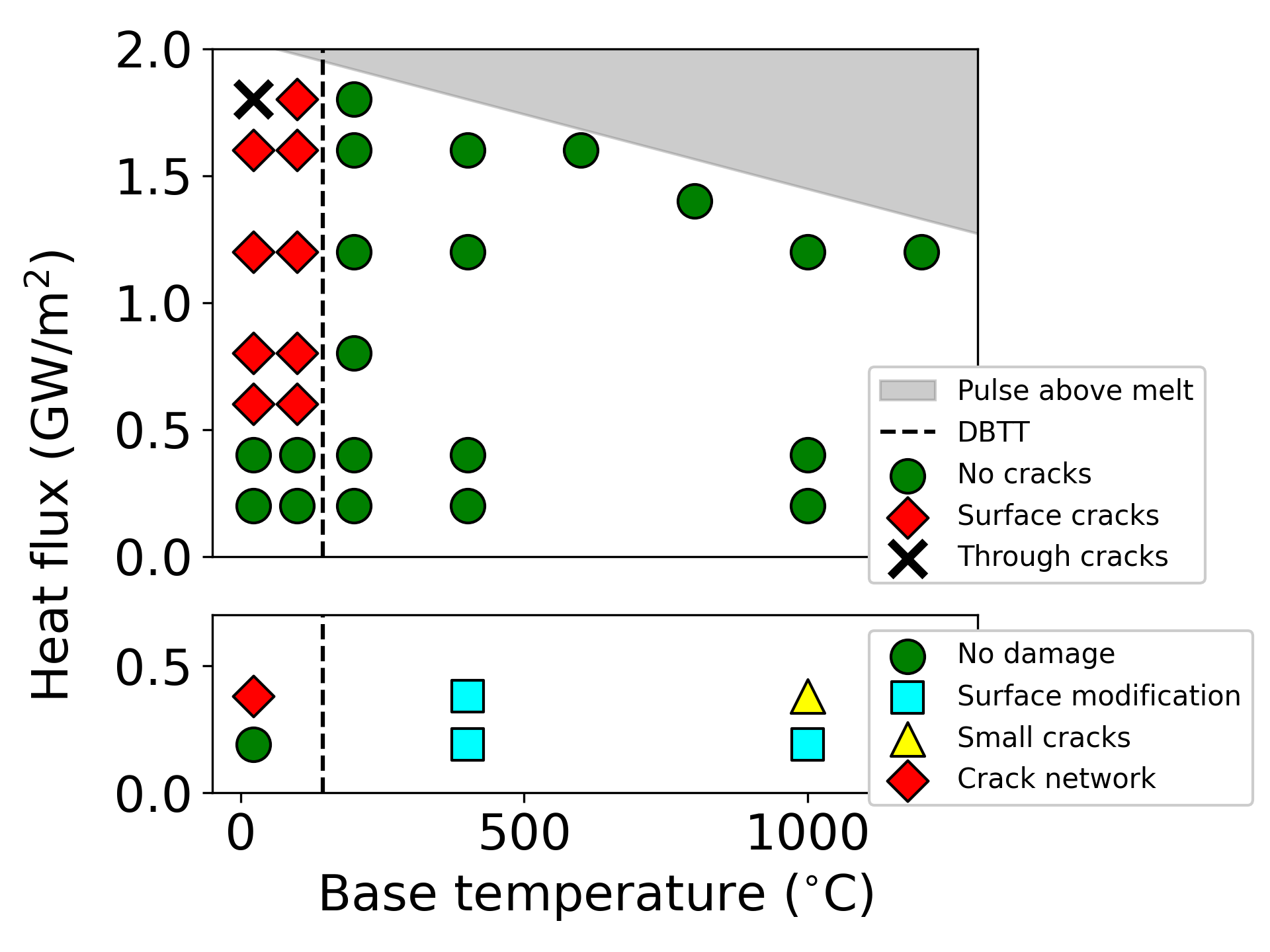}
    \caption{CabanaPD predictions (upper) and \PS{experimental results from}~\citet{wirtz2016} (lower) across heat flux and base temperature after 100 cycles.} 
    \label{fig:damage_map}
\end{figure}

\begin{figure}[h!]
    \centering
    \begin{subfigure}[b]{0.5\textwidth}
        \centering
        \includegraphics[width=\textwidth, trim=0cm 0cm 0cm 0cm, clip]{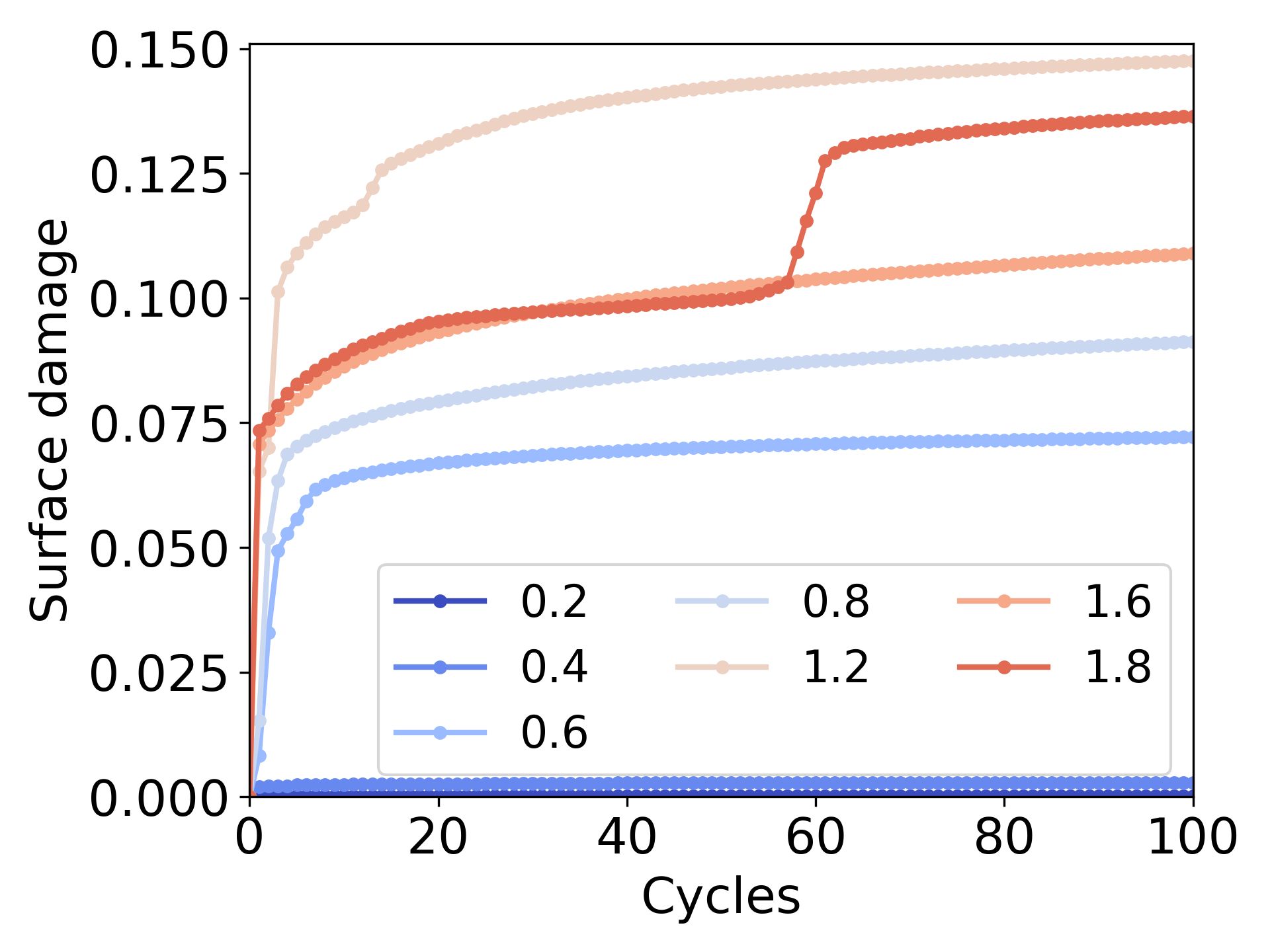}
        \subcaption{$T_0 = 23$$\degC$ }
        \label{fig:100cycles23}
    \end{subfigure}

    \begin{subfigure}[b]{0.5\textwidth}
        \centering
        \includegraphics[width=\textwidth]{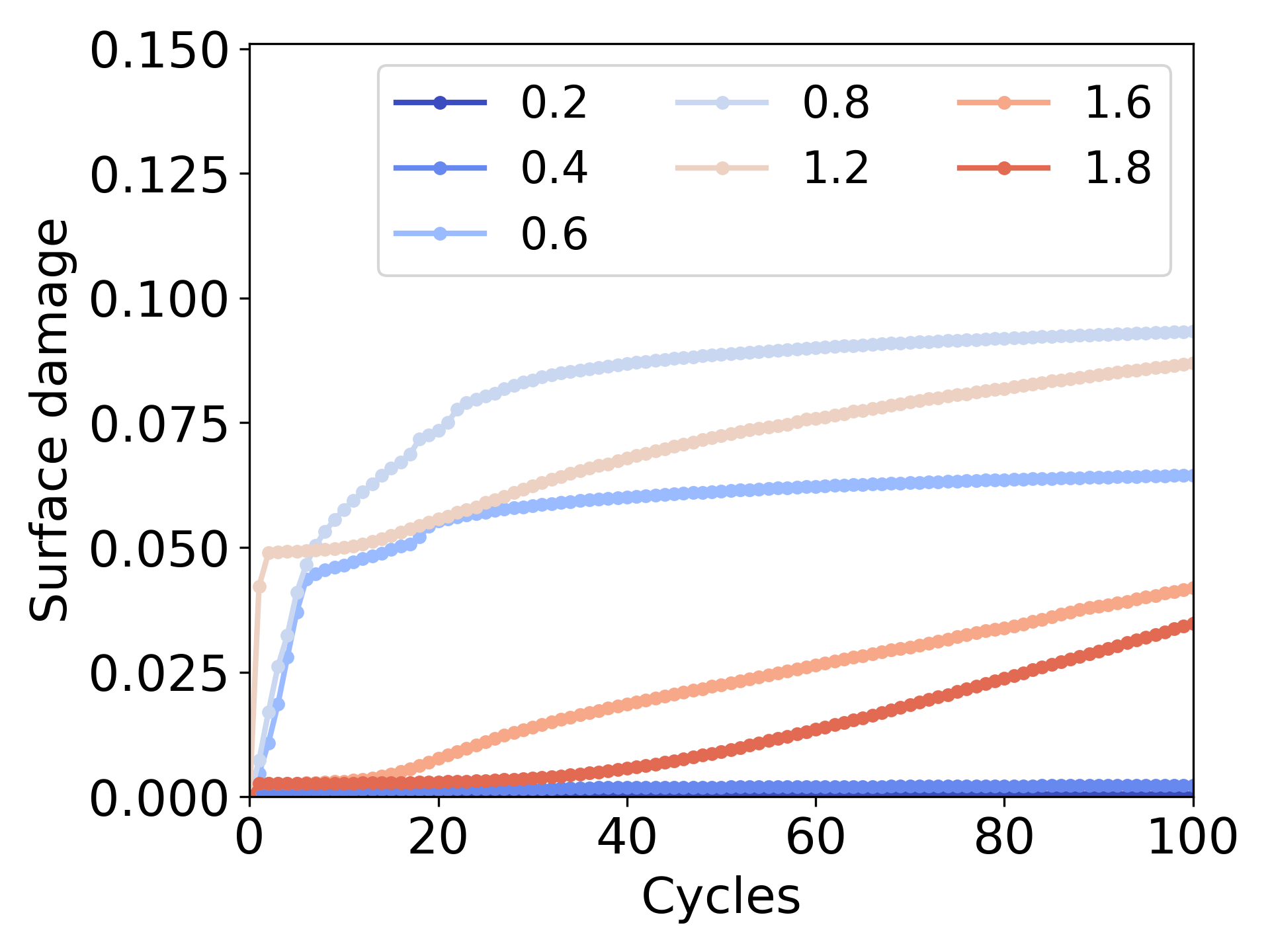}
        \subcaption{$T_0 = 100$$\degC$ }
        \label{fig:100cycles100}
    \end{subfigure}
    \caption{CabanaPD surface damage as a function of cycle number for different heat flux values %
    (given in the legend in GW/m$^2$) and base temperatures (noted in the subfigure captions).
    }
    \label{fig:100cycles}
\end{figure}

The experimental results from~\citet{wirtz2016} are replotted in the lower panel of Figure~\ref{fig:damage_map}, with results classified into four categories: no damage, surface modification, small cracks, and crack network. We note first that for the former two categories cracks were not observed, while for the latter two they were. In addition, the label ``no damage'' is distinct from the 
\PSSS{damage definition given in~\eqref{eq: total damage discrete}}
and qualitatively matches the ``no cracks'' label.
No cases were reported with through cracks given the relatively low heat flux conditions tested.
At the resolution considered here\JW{in}, and without directly including microstructural effects, the simulations \JW{can only} predict macroscopic cracking, which most closely corresponds to the experimental crack-network category. 
Higher-resolution simulations may be able  to predict small, isolated cracks or surface modification, but the small cracks identified in~\citet{wirtz2013} are on the same order as the grid spacing used here. \JW{As such, they} cannot be resolved in the present simulations.

Overall, the %
simulations are consistent with the experimental results reported in~\cite{wirtz2016}: 
no cracks at room temperature for a heat flux of $0.2$ GW/m$^2$, surface cracks at room temperature \PSSS{for} heat flux\SAM{es above $0.4$} GW/m$^2$, 
 and no significant damage at $400\degC$ or $1000\degC$. While the simulation case of 0.4 GW/m$^2$ at room temperature showed a very small amount of damage after 100 cycles (see Figure~\ref{fig:classify(b)}), it did not crack as in the experiment. However, we consider this level of accuracy reasonable given that the experimental material properties were used directly without any calibration and the microstructure was not explicitly included in the simulations.
In addition, the simulation results show reasonable trends beyond the experimentally tested conditions: 
failure 
\PSSS{under}  
many conditions below the \ac{DBTT}, 
(primarily) increasing damage with increasing heat flux, and decreasing damage with increasing base temperature. 
 
The CabanaPD simulations also provide significant detail on the evolution of surface damage and cracking as a function of cycle number. Representative cases are shown at both room temperature and 100$\degC$ in Figure~\ref{fig:100cycles} across heat flux conditions.
While the surface damage results generally increase with increasing heat flux, the variability and sensitivity of fracture patterns lead to variability in the surface damage that is commensurate with those effects. 
We also note that in the case of 1.8 GW/m$^2$ at room temperature, only part of the total damage is included in the surface damage in Figure~\ref{fig:100cycles} because that value is calculated only over the restricted region defined in Section~\ref{sec:sim}.
At 100$\degC$, there was less damage at each heat flux than at room temperature (with the exception of 0.8 GW/m$^2$ where they were nearly equal) and the damage evolved more slowly. 
For the two highest heat fluxes, 1.6 and 1.8 GW/m$^2$, 
the surface damage at 100$\degC$ was significantly lower than that at room temperature and at lower heat fluxes, a trend also observed in many experiments reported in~\citet{wirtz2013}.
These two conditions continued to evolve significant damage in subsequent cycles in contrast to the rest of the conditions tested.

We note that the highest surface damage observed in a system classified 
\PSSS{as having}
no cracks occurred at a heat flux of $0.4$ GW/m$^2$ at room temperature and remained below 0.003;
the maximum particle damage (ignoring the edge of the domain) was below 0.20. 
\SAM{This is indicative of this case being closer to cracking than the remaining conditions, as occurred in the experiment}~\citep{wirtz2016}.
Conversely, the lowest final surface damage for a system with surface cracks was above 0.03, \PSSS{i.e, more that 10$\times$ higher,} with maximum particle damage above 0.49, and occurred at a heat flux of 1.8 GW/m$^2$ and 100\PSSS{$\degC$}.
This highlights the large gap between conditions which did and did not crack.

In addition to the small difference\PSSS{, relative to the experiments,} in the minimum heat flux necessary to crack the sample at room temperature, other discrepancies are present in the results. Because the grain structure is not considered in this work, the intergranular fracture observed in the experimental results is not reproducible~\citep{wirtz2013, wirtz2016}. Instead, the samples generally fail at the location of the highest thermal gradient\PSSS{:} the edge of the heated region.
The application of the thermal boundary condition across multiple layers of particles below the surface, to avoid issues at the surface, may also contribute to the lack of cracking in the center of the sample in some cases. Finally, while the physics included here predict the 100-cycle cases well, the 1000-cycle cases also tested experimentally in~\cite{wirtz2016} may require plastic hardening and other cyclic damage effects.

\section{Conclusions}
\label{sec::conclusions}

In this paper, we developed a 
thermo-mechanical modeling and simulation capability 
based on \acf{PD} for tungsten \acfp{PFM} subjected to fusion-relevant e-beam heat fluxes. By implementing coupled thermo-mechanical and heat\PS{-}transfer models and solvers in the CabanaPD code, we successfully captured  experimentally observed trends in cracking and damage \PS{under} cyclic loading in recrystallized tungsten across \PS{a range of} base-temperature and heat-flux e-beam conditions. Our simulations enable detailed investigation of \PSSS{the} prediction of \PSSS{the} number of cycles \PSSS{required} to initiate surface cracking, as well as 
\PSSS{identification of conditions that lead to component-critical through-thickness cracks.}

The availability of experimental data is paramount for successful predictive simulation. The characterization of temperature-dependent material properties and the thermal-shock results reported by~\cite{wirtz2013} and~\cite{wirtz2016} are comprehensive and nearly complete; however, the lack of room-temperature data and detailed measurements of the ductile-to-brittle transition temperature (DBTT) remains a critical need for predicting damage in tungsten~\citep{yin_2021}.

Notably, the bond-based \ac{PD} modeling framework employed in this work was sufficient to achieve good agreement with experimental results. This framework incorporated a thermo-elasto-perfectly plastic \PS{mechanical} model coupled with nonlocal heat transfer, both \PS{using} temperature-dependent material properties that are essential for properly resolving the material response under thermal loading. 
However, several model improvements could be considered. First, a more general state-based \ac{PD} framework \PS{\citep{Silling2007}} could represent a broader range of deformation modes, for example \PS{by} eliminating the fixed Poisson's ratio  restriction inherent in bond-based \ac{PD} models~\citep{Silling2000}. 
Second, incorporating cycle-dependent material properties, hardening, and creep models could further improve the physical fidelity of the simulations and will likely be necessary for cases with higher numbers of thermal cycles, particularly up to millions of cycles~\citep{wirtz2017}. 
Third, 
calibration of the \ac{PD} model parameters will likely be necessary for more precise comparison with experiments, rather than direct adoption of experimentally measured material \PS{properties}, as was done here.
\JW{And fourth, the implementation of heat flux boundary conditions would more faithfully represent the experimental thermal loading.}
In addition, we note the limitations of the isotropic \ac{PD} model used here for recrystallized grain structures; anisotropic models would enable simulation of forged tungsten alloys commonly used as \acp{PFM}, whose longitudinal and transverse responses differ\PS{,} as reported in~\cite{wirtz2013}. 

There are many other possible improvements to consider for fusion-specific fracture predictions, most notably the incorporation of neutron flux and other plasma effects on the thermo-mechanical properties that contribute to failure. These effects have been investigated in previous work, but without explicit prediction of fracture~\citep{Durif2022}.
Direct connection to additional physics components, such as recrystallization and plasma effects on microstructure~\citep{yu2026}, should be established and linked to the current continuum-scale work. Notably, \ac{PD} modeling 
has also been demonstrated for explicit microstructural fracture prediction, resolving grain boundaries and enabling both %
\PS{intergranular and transgranular fracture} %
\PS{~\citep{GHAJARI2014431, LI2020167}}. 
This increased microstructural complexity 
will likely also require additional computational tools because experimentally characterized microstructures before and after exposure conditions remain limited.
Generative machine learning can be used to augment existing workflows and available experimental data, as previously demonstrated specifically for 
tungsten~\citep{RJ2025}, as well as more generally~\citep{fernandez-zelaia2024}.
In addition, the integration of experimental results with process-structure-property-performance predictions through digital twin approaches will become increasingly important, with success found in areas such as additive manufacturing~\citep{myna}.

Despite these limitations, 
the presented modeling and simulation framework is flexible and the results reported here lay a foundation for future enhancements, further experimental validation, and predictive modeling.
Overall, this work provides an approach for predicting fracture in \acp{PFM} under the extreme conditions encountered in fusion reactors.

\section*{Acknowledgments}
Research supported in part by the Laboratory Directed Research and Development Program of Oak Ridge National Laboratory, managed by UT-Battelle, LLC, for the U.S. Department of Energy,  
and by the U.S. Department of Energy, 
Office of Science, Office of Fusion Energy Sciences. 
This research used resources of the Oak Ridge Leadership Computing Facility at the Oak Ridge National Laboratory, which is supported by the Advanced Scientific Computing Research programs in the Office of Science of the U.S. Department of Energy under Contract No. DE-AC05-00OR22725. 

The authors thank Mohammed Alnaggar for valuable scientific discussions and for guidance on the machine-compliance correction; Stewart Silling for insightful guidance on implementing the microplastic peridynamic model; Jakob Bludau for implementation of the adaptive dynamic relaxation method; and Cory Hauck for reviewing the manu\-script, 
for helpful technical discussions, and for assistance with the Frontier allocation proposal.

\printcredits

\bibliographystyle{cas-model2-names}

\bibliography{references}

@article{Loewenhoff_2011,
doi = {10.1088/0031-8949/2011/T145/014057},
year = {2011},
month = {dec},
publisher = {},
volume = {2011},
number = {T145},
pages = {014057},
author = {Loewenhoff, {\relax Th} and Bürger, A and Linke, J and Pintsuk, G and Schmidt, A and Singheiser, L and Thomser, C},
title = {Evolution of tungsten degradation under combined high cycle edge-localized mode and steady-state heat loads},
journal = {Physica Scripta},
}

@article{Linsmeier_2017,
doi = {10.1088/1741-4326/aa6f71},
year = {2017},
month = {jun},
publisher = {IOP Publishing},
volume = {57},
number = {9},
pages = {092007},
author = {Linsmeier, Ch. and Rieth, M. and Aktaa, J. and Chikada, T. and Hoffmann, A. and Hoffmann, J. and Houben, A. and Kurishita, H. and Jin, X. and Li, M. and Litnovsky, A. and Matsuo, S. and von Müller, A. and Nikolic, V. and Palacios, T. and Pippan, R. and Qu, D. and Reiser, J. and Riesch, J. and Shikama, T. and Stieglitz, R. and Weber, T. and Wurster, S. and You, J.-H. and Zhou, Z.},
title = {Development of advanced high heat flux and plasma-facing materials},
journal = {Nuclear Fusion},
}

@article{Changjun2017,
author = {Changjun, Li and Zhu, Dahuan and Li, Xiangbin and Wang, Baoguo and Chen, Junling},
year = {2017},
month = {07},
pages = {68-73},
title = {Thermal–stress analysis on the crack formation of tungsten during fusion relevant transient heat loads},
volume = {13},
journal = {Nuclear Materials and Energy},
doi = {10.1016/j.nme.2017.06.008}
}

@article{Durif2022,
title = {{T-REX}: Numerical tool for tungsten damage assessment for {DEMO}},
journal = {Journal of Nuclear Materials},
volume = {569},
pages = {153906},
year = {2022},
issn = {0022-3115},
doi = {10.1016/j.jnucmat.2022.153906},
author = {A. Durif and M. Richou and J.-M. Bergheau and L. Gallais and G. Kermouche and G. Pintsuk},
}

@article{Durif2023,
title = {Edge cracking of {WEST} tungsten actively cooled plasma facing components after plasma operation},
journal = {Fusion Engineering and Design},
volume = {188},
pages = {113441},
year = {2023},
issn = {0920-3796},
doi = {10.1016/j.fusengdes.2023.113441},
author = {A. Durif and M. Richou and J.-M. Bergheau and Y. Corre and M. Diez and P. Reilhac and J.P. Gunn and E. Tsitrone},
}

@book{wirtz2013,
  title={Thermal shock behaviour of different tungsten grades under varying conditions},
  author={Wirtz, Oliver Marius},
  volume={161},
  year={2013},
  publisher={Forschungszentrum J{\"u}lich}
}

@article{RJ2025,
title = {Machine learning for microstructure synthesis in fusion materials: A physics-aware validation framework for tungsten plasma-facing components},
journal = {Journal of Nuclear Materials},
volume = {623},
pages = {156473},
year = {2026},
issn = {0022-3115},
doi = {10.1016/j.jnucmat.2026.156473},
author = {R. Juneja and M. Wirtz and J. W. Coenen and J. Rapp and E. A. Unterberg},
}

@article{wirtz2016,
	title = {Thermal shock tests to qualify different tungsten grades as plasma facing material},
	volume = {T167},
	issn = {0031-8949, 1402-4896},
	doi = {10.1088/0031-8949/T167/1/014015},
	language = {en},
	urldate = {2026-01-12},
	journal = {Physica Scripta},
	author = {Wirtz, M and Linke, J and Loewenhoff, {\relax Th} and Pintsuk, G and Uytdenhouwen, I},
	month = feb,
	year = {2016},
	pages = {014015},
}

@misc{cabanapd,
  author       = {Sam Reeve and
                  Pablo Seleson and
                  Jakob Bludau and
                  Irán Shepherd and
                  Austin Isner and
                  Kinan Bezem},
  title        = {{CabanaPD}: Version 0.5},
  month        = may,
  year         = 2026,
  publisher    = {Zenodo},
  version      = {0.5.0},
  doi          = {10.5281/zenodo.20449362},
}

@article{Silling2000,
title = {Reformulation of elasticity theory for discontinuities and long-range forces},
journal = {Journal of the Mechanics and Physics of Solids},
volume = {48},
number = {1},
pages = {175-209},
year = {2000},
issn = {0022-5096},
doi = {10.1016/S0022-5096(99)00029-0},
author = {S. A. Silling},
}

@article{SILLING2005,
title = {A meshfree method based on the peridynamic model of solid mechanics},
journal = {Computers $\&$ Structures},
volume = {83},
number = {17},
pages = {1526-1535},
year = {2005},
issn = {0045-7949},
doi = {10.1016/j.compstruc.2004.11.026},
author = {S.A. Silling and E. Askari},
}

@ARTICLE{CoPA-paper,
  author={Reeve, Samuel Temple and Fattebert, Jean-Luc and DeWitt, Stephen and Seleson, Pablo and Joy, David and Slattery, Stuart and Scheinberg, Aaron and Halver, Rene and Junghans, Christoph and Negre, Christian F. A. and Wall, Michael E. and Zhang, Yu and Niklasson, Anders M. and Belak, James and Mniszewski, Susan M. and Perez, Danny},
  journal={Computing in Science $\&$ Engineering}, 
  title={Co-design for Particle Applications at Exascale}, 
  year={2024},
  volume={26},
  number={2},
  pages={43-52},
  doi={10.1109/MCSE.2024.3384052}}

@article{BOBARU2010,
title = {The peridynamic formulation for transient heat conduction},
journal = {International Journal of Heat and Mass Transfer},
volume = {53},
number = {19},
pages = {4047-4059},
year = {2010},
issn = {0017-9310},
doi = {10.1016/j.ijheatmasstransfer.2010.05.024},
author = {Florin Bobaru and Monchai Duangpanya},
}

@article{MACEK2007,
title = {Peridynamics via finite element analysis},
journal = {Finite Elements in Analysis and Design},
volume = {43},
number = {15},
pages = {1169-1178},
year = {2007},
issn = {0168-874X},
doi = {10.1016/j.finel.2007.08.012},
author = {Richard W. Macek and Stewart A. Silling},
}

@article{GIANNAKEAS20183037,
title = {Simulation of thermal shock cracking in ceramics using bond-based peridynamics and {FEM}},
journal = {Journal of the European Ceramic Society},
volume = {38},
number = {8},
pages = {3037-3048},
year = {2018},
issn = {0955-2219},
doi = {10.1016/j.jeurceramsoc.2017.12.039},
author = {Ilias N. Giannakeas and Theodosios K. Papathanasiou and Hamid Bahai},
}

@article{DANTUONO201774,
title = {Thermal shock response via weakly coupled peridynamic thermo-mechanics},
journal = {International Journal of Solids and Structures},
volume = {129},
pages = {74-89},
year = {2017},
issn = {0020-7683},
doi = {10.1016/j.ijsolstr.2017.09.010},
author = {Pietro D’Antuono and Marco Morandini},
}

@Article{Silling2007,
author={Silling, S. A.
and Epton, M.
and Weckner, O.
and Xu, J.
and Askari, E.},
title={Peridynamic States and Constitutive Modeling},
journal={Journal of Elasticity},
year={2007},
month={Aug},
day={01},
volume={88},
number={2},
pages={151-184},
issn={1573-2681},
doi={10.1007/s10659-007-9125-1},
}

@Article{Ha2010,
author={Ha, Youn Doh
and Bobaru, Florin},
title={Studies of dynamic crack propagation and crack branching with peridynamics},
journal={International Journal of Fracture},
year={2010},
month={Mar},
day={01},
volume={162},
number={1},
pages={229-244},
issn={1573-2673},
doi={10.1007/s10704-010-9442-4},
}

@Article{Silling2010,
author={Silling, S. A.
and Weckner, O.
and Askari, E.
and Bobaru, F.},
title={Crack nucleation in a peridynamic solid},
journal={International Journal of Fracture},
year={2010},
month={Mar},
day={01},
volume={162},
number={1},
pages={219-227},
doi={10.1007/s10704-010-9447-z},
}

@article{kokkos,
author = {Trott, Christian R. and Lebrun-Grandié, Damien and Arndt, Daniel and Ciesko, Jan and Dang, Vinh and Ellingwood, Nathan and Gayatri, Rahulkumar and Harvey, Evan and Hollman, Daisy S. and Ibanez, Dan and Liber, Nevin and Madsen, Jonathan and Miles, Jeff and Poliakoff, David and Powell, Amy and Rajamanickam, Sivasankaran and Simberg, Mikael and Sunderland, Dan and Turcksin, Bruno and Wilke, Jeremiah},
doi = {10.1109/TPDS.2021.3097283},
journal = {IEEE Transactions on Parallel and Distributed Systems},
number = {4},
pages = {805--817},
title = {{Kokkos 3: Programming Model Extensions for the Exascale Era}},
volume = {33},
year = {2022}
}

@article{cabana,
  doi = {10.21105/joss.04115},
  year = {2022},
  publisher = {The Open Journal},
  volume = {7},
  number = {72},
  pages = {4115},
  author = {Stuart Slattery and Samuel Temple Reeve and Christoph Junghans and Damien Lebrun-Grandié and Robert Bird and Guangye Chen and Shane Fogerty and Yuxing Qiu and Stephan Schulz and Aaron Scheinberg and Austin Isner and Kwitae Chong and Stan Moore and Timothy Germann and James Belak and Susan Mniszewski},
  title = {Cabana: A Performance Portable Library for Particle-Based Simulations},
  journal = {Journal of Open Source Software}
}

@article{yin_2021,
	title = {Ductile to brittle transition temperature of advanced tungsten alloys for nuclear fusion applications deduced by miniaturized three-point bending tests},
	volume = {95},
	issn = {0263-4368},
	doi = {10.1016/j.ijrmhm.2020.105464},
	urldate = {2026-06-11},
	journal = {International Journal of Refractory Metals and Hard Materials},
	author = {Yin, Chao and Terentyev, Dmitry and Zhang, Tao and Nogami, Shuhei and Antusch, Steffen and Chang, Chih-Cheng and Petrov, Roumen H. and Pardoen, Thomas},
	month = feb,
	year = {2021},
	pages = {105464},
}

@article{wang2018,
        title = {A 3-{D} conjugated bond-pair-based peridynamic formulation for initiation and propagation of cracks in brittle solids},
        volume = {134},
        issn = {0020-7683},
        doi = {10.1016/j.ijsolstr.2017.10.022},
        urldate = {2026-03-25},
        journal = {International Journal of Solids and Structures},
        author = {Wang, Yunteng and Zhou, Xiaoping and Wang, Yuan and Shou, Yundong},
        month = mar,
        year = {2018},
        pages = {89--115},
}

@article{yu2026,
title = {Irradiation-driven recrystallization in fusion-grade tungsten: A mesoscale, microstructure-aware model},
journal = {Acta Materialia},
volume = {320},
pages = {122703},
year = {2026},
issn = {1359-6454},
doi = {10.1016/j.actamat.2026.122703},
author = {Jinxin Yu and Sicong He and Giacomo Po and Jason R. Trelewicz and Timothy J. Rupert and Jaime Marian},
}

@article{fernandez-zelaia2024,
	title = {Digital polycrystalline microstructure generation using diffusion probabilistic models},
	volume = {33},
	issn = {2589-1529},
	doi = {10.1016/j.mtla.2023.101976},
	urldate = {2026-08-03},
	journal = {Materialia},
	author = {Fernandez-Zelaia, Patxi and Cheng, Jiahao and Mayeur, Jason and Ziabari, Amir Koushyar and Kirka, Michael M.},
	month = mar,
	year = {2024},
	pages = {101976},
}

@article{li2015,
	title = {Experimental and computational study of damage behavior of tungsten under high energy electron beam irradiation},
	volume = {135},
	issn = {0013-7944},
	doi = {10.1016/j.engfracmech.2015.01.017},
	urldate = {2026-08-04},
	journal = {Engineering Fracture Mechanics},
	author = {Li, Muyuan and Sommerer, Mathias and Werner, Ewald and Lampenscherf, Stefan and Steinkopff, Thorsten and Wolfrum, Philipp and You, Jeong-Ha},
	month = feb,
	year = {2015},
	pages = {64--80},
}

@article{chen2025,
	title = {Computational study of tungsten cracking propagation under {ELM}-like high heat flux conditions},
	volume = {605},
	issn = {0022-3115},
	doi = {10.1016/j.jnucmat.2024.155567},
	urldate = {2026-08-04},
	journal = {Journal of Nuclear Materials},
	author = {Chen, Li and Jiang, Shi and Zhang, Shilong and Li, Weihao and Hao, Shiyu and Li, Chengcheng and Shi, Huantong and Li, Xingwen},
	month = feb,
	year = {2025},
	pages = {155567},
}

@Article{Trageser2020,
author={Trageser, Jeremy and Seleson, Pablo},
title={Bond-Based Peridynamics: a Tale of Two {Poisson's} Ratios},
journal={Journal of Peridynamics and Nonlocal Modeling},
year={2020},
volume={2},
number={3},
pages={278-288},
doi={10.1007/s42102-019-00021-x},
}

@article{BOBARU20122764,
title = {A peridynamic formulation for transient heat conduction in bodies with evolving discontinuities},
journal = {Journal of Computational Physics},
volume = {231},
number = {7},
pages = {2764-2785},
year = {2012},
issn = {0021-9991},
doi = {10.1016/j.jcp.2011.12.017},
author = {Florin Bobaru and Monchai Duangpanya},
}

@article{wirtz2017,
  title={Transient heat load challenges for plasma-facing materials during long-term operation},
  author={Wirtz, Marius and Linke, J and Loewenhoff, {\relax Th} and Pintsuk, G and Uytdenhouwen, I},
  journal={Nuclear Materials and Energy},
  volume={12},
  pages={148--155},
  year={2017},
  doi = {10.1016/j.nme.2016.12.024},
  publisher={Elsevier}
}

@article{BelytschkoBlack1999,
author = {Belytschko, T. and Black, T.},
title = {Elastic crack growth in finite elements with minimal remeshing},
journal = {International Journal for Numerical Methods in Engineering},
volume = {45},
number = {5},
pages = {601-620},
doi = {10.1002/(SICI)1097-0207(19990620)45:5<601::AID-NME598>3.0.CO;2-S},
year = {1999}
}

@article{MoësDolbowBelytschko-1999,
author = {Moës, Nicolas and Dolbow, John and Belytschko, Ted},
title = {A finite element method for crack growth without remeshing},
journal = {International Journal for Numerical Methods in Engineering},
volume = {46},
number = {1},
pages = {131-150},
doi = {10.1002/(SICI)1097-0207(19990910)46:1<131::AID-NME726>3.0.CO;2-J},
year = {1999}
}

@article{XuNeedleman1994,
title = {Numerical simulations of fast crack growth in brittle solids},
journal = {Journal of the Mechanics and Physics of Solids},
volume = {42},
number = {9},
pages = {1397-1434},
year = {1994},
issn = {0022-5096},
doi = {10.1016/0022-5096(94)90003-5},
author = {X.-P. Xu and A. Needleman},
}

@article{SwensonIngraffea1988,
  author  = {Swenson, D. V. and Ingraffea, A. R.},
  title   = {Modeling mixed-mode dynamic crack propagation using finite elements: Theory and applications},
  journal = {Computational Mechanics},
  year    = {1988},
  volume  = {3},
  number  = {6},
  pages   = {381--397},
  doi     = {10.1007/BF00301139},
}

@article{ZU20242349,
title = {An Elastoplastic Fracture Model Based on Bond-Based Peridynamics},
journal = {CMES - Computer Modeling in Engineering and Sciences},
volume = {140},
number = {3},
pages = {2349-2371},
year = {2024},
issn = {1526-1492},
doi = {10.32604/cmes.2024.050488},
author = {Liping Zu and Yaxun Liu and Haoran Zhang and Lisheng Liu and Xin Lai and Hai Mei},
}

@article{linke2019,
  title={Challenges for plasma-facing components in nuclear fusion},
  author={Linke, Jochen and Du, Juan and Loewenhoff, {\relax Th} and Pintsuk, Gerald and Spilker, Benjamin and Steudel, Isabel and Wirtz, Marius},
  journal={Matter and Radiation at Extremes},
  volume={4},
  number={5},
  year={2019},
  pages={056201},
  publisher={AIP Publishing},
  doi={10.1063/1.5090100}
}

@article{myna,
  title={Myna: Connecting powder bed fusion build data to simulation tools for digital twin applications},
  author={Knapp, Gerald L and Reeve, Samuel Temple and Coleman, John and Rolchigo, Matthew and Stump, Benjamin and DeWitt, Stephen and Plotkowski, Alex},
  journal={Computational Materials Science},
  volume={258},
  pages={114094},
  year={2025},
  doi = {10.1016/j.commatsci.2025.114094},
  publisher={Elsevier}
}

@article{GHAJARI2014431,
title = {A peridynamic material model for the analysis of dynamic crack propagation in orthotropic media},
journal = {Computer Methods in Applied Mechanics and Engineering},
volume = {276},
pages = {431-452},
year = {2014},
issn = {0045-7825},
doi = {10.1016/j.cma.2014.04.002},
author = {M. Ghajari and L. Iannucci and P. Curtis},
}

@article{LI2020167,
title = {Thermally-induced fracture analysis of polycrystalline materials by using peridynamics},
journal = {Engineering Analysis with Boundary Elements},
volume = {117},
pages = {167-187},
year = {2020},
issn = {0955-7997},
doi = {10.1016/j.enganabound.2020.04.016},
author = {Mingyang Li and Wei Lu and Erkan Oterkus and Selda Oterkus},
}

\end{document}